%% file: autosam.tex
\documentclass[twocolumn]{autart}    

\usepackage{siunitx}
\usepackage{textcomp}
\usepackage{eurosym}
\usepackage{rotating}
\usepackage{tikz}
\usepackage{graphicx}          
\input{sections/common/preamble}
\usepackage[square, numbers]{natbib}
\usepackage{hyperref}
\begin{document}
\input{sections/common/sym_defs}

\begin{frontmatter}

\title{Long-term experimental study of price responsive predictive control in a real occupied single-family house with heat pump\thanksref{footnoteinfo}}

\thanks[footnoteinfo]{This work is funded the Danish Energy Agency through the EUDP project OPSYS 2.0 (Case num.: 64018-0581) and the Department of Electronic Systems at Aalborg University.}

\author[aau]{Simon Thorsteinsson}\ead{sith@es.aau.dk},    
\author[neogrid]{Alex Arash Sand Kalaee}\ead{ask@neogrid.dk},
\author[neogrid]{Pierre Vogler-Finck}\ead{pvf@neogrid.dk},
\author[neogrid]{Henrik Lund St\ae rmose}\ead{hls@neogrid.dk},
\author[dti]{Ivan Katic}\ead{ik@teknologisk.dk},
\author[aau]{Jan Dimon Bendtsen}\ead{dimon@es.aau.dk}

\address[aau]{Section for Automation and Control, Aalborg University, Aalborg, Denmark}  
\address[neogrid]{Neogrid Technologies ApS, Aalborg, Denmark}  
\address[dti]{Danish Technological Institute, Taastrup, Denmark}

\begin{keyword}                           
Hierarchical model predictive control; Heat pump; Retrofit building control; Real experiment; Load shifting; Demand side management.
\end{keyword}                               

\begin{abstract}
The continuous introduction of renewable electricity and increased consumption through electrification of the transport and heating sector challenges grid stability. This study investigates \loadshifting\ through \dsm\ as a solution. We present a four-month experimental study of a low-complexity, hierarchical Model Predictive Control approach for \dsm\ in a near-zero emission occupied single-family house in Denmark. The control algorithm uses a price signal, weather forecast, a single-zone building model, and a non-linear heat pump efficiency model to generate a \spaceHeating\ schedule. The weather-compensated, commercial heat pump is made to act \smartGrid-ready through outdoor temperature input override to enable heat boosting and forced stops to accommodate the heating schedule. The cost reduction from the controller ranged from \comfThreePercSaveNoDec-\comfFourPercSaveNoDec\% depending on the chosen comfort level. The experiment demonstrates that \loadshifting\ is feasible and cost-effective, even without energy storage, and that the current price scheme provides an incentive for Danish end-consumers to shift heating loads. However, issues related to controlling the heat pump through input-manipulation were identified, and the authors propose a more promising path forward involving coordination with manufacturers and regulators to make commercial heat pumps truly smart grid-ready.
\end{abstract}

\end{frontmatter}

\input{sections/introduction/introduction}
\input{sections/method/method}
\input{sections/results/results}
\input{sections/conclusion/conclusion}
\input{sections/admin/admin}
\begin{ack}                               
The authors would like to express their sincere gratitude to the family for risking money and comfort by letting us carry out this experiment on their house. This work is funded the Danish Energy Agency through the EUDP project OPSYS 2.0 (Case num.: 64018-0581).  
\end{ack}






\bibliographystyle{plainnat}

\setcitestyle{numbers}
\bibliography{autosam,mpc_dh1,software}           
\appendix
\input{sections/appendix/appendix}





\end{document}

%% file: sections/common/preamble.tex
\usepackage{amsmath}
\usepackage{amssymb}
\usepackage[framed,amsmath,thmmarks]{ntheorem}

\usepackage{graphicx}
\usepackage{float}
\usepackage{stfloats}
\usepackage{tabularx}
\usepackage{booktabs}
\usepackage{todonotes}
\usepackage{siunitx}

\usepackage[geometry]{ifsym}

\renewcommand{\arraystretch}{1.1}

\definecolor{ref1}{rgb}{0.12156862745098039, 0.4666666666666667, 0.7058823529411765}
\definecolor{ref2}{rgb}{0.7372549019607844, 0.7411764705882353, 0.13333333333333333}
\definecolor{ref3}{rgb}{0.8392156862745098, 0.15294117647058825, 0.1568627450980392}
\definecolor{ref4}{rgb}{1.0, 0.4980392156862745, 0.054901960784313725}

\newcommand{\figref}[1]{Fig.~\ref{#1}}



%% file: sections/common/sym_defs.tex
\newcommand{\perk}[1]{\ensuremath{\{k#1\}}}
\newcommand{\samk}[1]{\ensuremath{[k#1]}}
\newcommand{\timt}[1]{\ensuremath{(t#1)}}

\newcommand{\T}{\ensuremath{T}}             
\newcommand{\dotT}{\ensuremath{\dot{\T}}}   
\newcommand{\Pre}{\ensuremath{p}}           
\newcommand{\pow}{\ensuremath{P}}           
\newcommand{\energy}{\ensuremath{E}}        
\newcommand{\perc}{\ensuremath{\alpha}}     
\newcommand{\qprop}{\ensuremath{\alpha}}
\newcommand{\nomqprop}{\ensuremath{\overline{\qprop}}}
\newcommand{\U}{\ensuremath{U}}             %
\newcommand{\Capa}{\ensuremath{C}}
\newcommand{\Q}{\ensuremath{Q}}
\newcommand{\dotQ}{\dot{\Q}}
\newcommand{\vol}{\ensuremath{V}}
\newcommand{\dotV}{\dot{\V}}
\newcommand{\val}{\ensuremath{v}}
\newcommand{\Area}{\ensuremath{A}}
\newcommand{\mass}{\ensuremath{m}}
\newcommand{\dotm}{\ensuremath{q}}
\newcommand{\nomdotm}{\overline{\dotm}}
\newcommand{\nomflow}{\overline{\flow}}
\newcommand{\res}{\ensuremath{R}}			
\newcommand{\air}{\ensuremath{\text{a}}}
\newcommand{\spu}{\ensuremath{u}}			
\newcommand{\price}{\ensuremath{c}}
\newcommand{\dt}{\ensuremath{\Delta t}}
\newcommand{\len}{\ensuremath{\ell}}
\newcommand{\relative}{\ensuremath{\text{rel}}}
\newcommand{\vel}{\ensuremath{v}}           
\newcommand{\maxi}{\text{max}}
\newcommand{\Irr}{\ensuremath{I}}
\newcommand{\deltahp}{\ensuremath{\delta_{\hp}}}
\newcommand{\deltahpi}[1]{\ensuremath{\delta_{\hp,#1}}}
\newcommand{\costRoom}{\Rm}
\newcommand{\tsample}{\ensuremath{t_s}}

\newcommand{\xss}{\ensuremath{x}}			
\newcommand{\uss}{\ensuremath{u}}			
\newcommand{\fss}{\ensuremath{\tilde{d}}}			
\newcommand{\dss}{\ensuremath{\hat{d}}}

\newcommand{\room}{\ensuremath{\text{r}}} 			
\newcommand{\floor}{\ensuremath{\text{f}}}			

\newcommand{\brine}{\ensuremath{\text{B}}}			
\newcommand{\forward}{\ensuremath{\text{F}}}		
\newcommand{\return}{\ensuremath{\text{R}}}			
\newcommand{\fh}{\ensuremath{\text{FH}}}			
\newcommand{\hp}{\ensuremath{\text{HP}}}			
\newcommand{\area}{\ensuremath{\text{A}}}			
\newcommand{\water}{\ensuremath{\text{w}}}			
\newcommand{\hot}{\ensuremath{\text{H}}}			
\newcommand{\cold}{\ensuremath{\text{C}}}			
\newcommand{\tank}{\ensuremath{\text{ta}}}			
\newcommand{\pv}{\ensuremath{\text{PV}}}			%
\newcommand{\sensor}{\ensuremath{\text{s}}}			
\newcommand{\envel}{\ensuremath{\text{e}}}			
\newcommand{\cop}{\ensuremath{\text{COP}}}			
\newcommand{\comp}{\ensuremath{\text{com}}}			
\newcommand{\evap}{\ensuremath{\text{eva}}}			
\newcommand{\source}{\ensuremath{\text{src}}}		
\newcommand{\coTwo}{\ensuremath{\text{CO}_2}}
\newcommand{\coTwoTax}{\coTwo-tax}
\newcommand{\pipe}{\ensuremath{\text{p}}}
\newcommand{\reference}{\ensuremath{\text{ref}}}
\newcommand{\grid}{\ensuremath{\text{G}}}
\newcommand{\diag}{\ensuremath{\text{diag}}}
\newcommand{\bess}{\ensuremath{\text{b}}}
\newcommand{\comf}{\ensuremath{\text{cmf}}}
\newcommand{\expo}{\ensuremath{\text{EX}}}
\newcommand{\impo}{\ensuremath{\text{IM}}}
\newcommand{\up}{\ensuremath{\text{up}}}
\newcommand{\dn}{\ensuremath{\text{dn}}}
\newcommand{\app}{\ensuremath{\text{APP}}}
\newcommand{\charge}{\ensuremath{\text{c}}}
\newcommand{\transport}{\ensuremath{\text{tr}}}
\newcommand{\partition}{\ensuremath{\text{part}}}
\newcommand{\consumption}{\ensuremath{\text{con}}}
\newcommand{\selfCons}{\ensuremath{\text{self}}}
\newcommand{\spot}{\ensuremath{\text{spot}}}
\newcommand{\tariff}{\ensuremath{\text{tariff}}}
\newcommand{\daily}{\ensuremath{\text{day}}}
\newcommand{\elec}{\ensuremath{\text{E}}}
\newcommand{\effic}{\ensuremath{\eta}}
\newcommand{\slack}{\ensuremath{\text{s}}}
\newcommand{\slackVar}{\ensuremath{S}}
\newcommand{\direct}{\ensuremath{\text{dir}}}
\newcommand{\Nslack}{\ensuremath{N_\slack}}
\newcommand{\Ncomf}{\ensuremath{N_\comf}}
\newcommand{\Nroom}{\ensuremath{N_\room}}
\newcommand{\TSO}{\ensuremath{\text{tso}}}
\newcommand{\VAT}{\ensuremath{\text{vat}}}
\newcommand{\experiment}{\ensuremath{\text{exp}}}
\newcommand{\compare}{\ensuremath{\text{cmp}}}
\newcommand{\cost}{\ensuremath{\text{cost}}}
\newcommand{\Ehp}{\ensuremath{\energy_{\hp}}}
\newcommand{\Ehpexp}{\ensuremath{\energy_{\hp, \experiment}}}
\newcommand{\Ehpcmp}{\ensuremath{\energy_{\hp, \compare}}}
\newcommand{\Epv}{\ensuremath{\energy_{\pv}}}
\newcommand{\Epvi}[1]{\ensuremath{\energy_{\pv,#1}}}
\newcommand{\Eg}{\ensuremath{\energy_{\grid}}}
\newcommand{\Econ}{\ensuremath{\energy_{\consumption}}}
\newcommand{\Eapp}{\ensuremath{\energy_{\app}}}
\newcommand{\Eself}{\ensuremath{\energy_{\selfCons}}}
\newcommand{\Eexp}{\ensuremath{\energy_{\expo}}}
\newcommand{\Eimp}{\ensuremath{\energy_{\impo}}}
\newcommand{\dailyExp}{\ensuremath{\text{day}_{\experiment}}}
\newcommand{\dailyCmp}{\ensuremath{\text{day}_{\compare}}}
\newcommand{\EpvCmp}{\Epvi{\compare}}
\newcommand{\EpvExp}{\Epvi{\experiment}}
\newcommand{\kf}{\ensuremath{\text{Kalman Filter}}}
\newcommand{\ukf}{\ensuremath{\text{Unscented }\kf}}
\newcommand{\Qref}{\ensuremath{\Q_{\reference}}}
\newcommand{\setVal}{\ensuremath{\mathcal{V}}}

\newcommand{\vval}{\ensuremath{{v}}}			

\newcommand{\pperA}{\ensuremath{g_\pipe}}
\newcommand{\dtr}{\ensuremath{\dt_\room}}
\newcommand{\dtri}[1]{\ensuremath{\dt_{\room,#1}}}
\newcommand{\dtest}{\ensuremath{\hat{\dt}}}
\newcommand{\dtrest}{\ensuremath{\dtest_\room}}
\newcommand{\dtrestj}[1]{\ensuremath{\dtest_{\room,#1}}}
\newcommand{\dttr}{\ensuremath{\dt_\transport}}
\newcommand{\dttrest}{\ensuremath{\dtest_\transport}} 
\newcommand{\dttri}[1]{\ensuremath{\dt_{\transport,#1}}}
\newcommand{\pperAavg}{\bar{\pperA}}

\newcommand{\lentr}{\len_{\transport}}
\newcommand{\lenR}{\len_{\return}}
\newcommand{\lentrest}{\hat{\len}_{\transport}}
\newcommand{\lentrestj}[1]{\hat{\len}_{\transport,#1}}
\newcommand{\lentrj}[1]{\len_{\transport,#1}}
\newcommand{\lenpart}{\len_\text{part}}
\newcommand{\amb}{\ensuremath{\text{a}}}			
\newcommand{\inter}{\ensuremath{\text{int}}}			
\newcommand{\occ}{\ensuremath{\text{occ}}}			
\newcommand{\loss}{\ensuremath{\text{loss}}}		
\newcommand{\SUN}{\ensuremath{\text{s}}}			
\newcommand{\sun}{\ensuremath{\text{\SUN}}}
\newcommand{\cloud}{\ensuremath{\text{cloud}}}

\newcommand{\IN}{\ensuremath{\text{in}}}			
\newcommand{\OUT}{\ensuremath{\text{out}}}			

\newcommand{\subjectTo}{\ensuremath{\text{s.t.}}}   
\newcommand{\priceElec}{\ensuremath{\price_\elec}}
\newcommand{\priceElecBuy}{\ensuremath{\price_\elec^{+}}}
\newcommand{\priceElecBuySeti}[1]{\ensuremath{\price_{\elec}^{+,#1}}}
\newcommand{\priceElecSell}{\ensuremath{\price_\elec^{-}}}
\newcommand{\priceComf}{\ensuremath{\price_\comf}}
\newcommand{\priceComfi}[1]{\ensuremath{\price_{\comf,#1}}}
\newcommand{\priceSlack}{\ensuremath{\price_\slack}}
\newcommand{\priceCoTwo}{\ensuremath{\price_{\coTwo}}}
\newcommand{\priceSpot}{\ensuremath{\price_{\spot}}}
\newcommand{\priceTariff}{\ensuremath{\price_{\tariff}}}
\newcommand{\priceTransport}{\ensuremath{\price_{\TSO}}}
\newcommand{\priceVat}{\ensuremath{\price_{\VAT}}}
\newcommand{\priceElecBuyi}[1]{\ensuremath{\price_{\elec,#1}^{+}}}
\newcommand{\priceElecBuyExp}{\priceElecBuyi{\experiment}}
\newcommand{\priceElecBuyCmp}{\priceElecBuyi{\compare}}
\newcommand{\costElecCmp}{\ensuremath{\cost_{\compare}}}
\newcommand{\costElecExp}{\ensuremath{\cost_{\experiment}}}
\newcommand{\deltaOp}{\ensuremath{\delta_{\op}}}

\newcommand{\priceEleci}[1]{\ensuremath{\price_{\elec,#1}}}

\newcommand{\Ar}{\A_\room}
\newcommand{\Vr}{\V_\room}
\newcommand{\Tr}{\ensuremath{\T_{\room}}}
\newcommand{\Tcold}{\ensuremath{\T_{\cold}}}
\newcommand{\Thot}{\ensuremath{\T_{\hot}}}
\newcommand{\flow}{\ensuremath{q}}
\newcommand{\Tref}{\ensuremath{\T_{\room,\reference}}}
\newcommand{\TFref}{\ensuremath{\T_{\forward,\reference}}}
\newcommand{\dotTr}{\ensuremath{\dot{\T}_{\room}}}
\newcommand{\Tf}{\ensuremath{\T_{\floor}}}
\newcommand{\dotTf}{\ensuremath{\dot{\T}_{\floor}}}
\newcommand{\Tw}{\ensuremath{\T_{\water}}}
\newcommand{\dotTw}{\ensuremath{\dot{\T}_{\water}}}
\newcommand{\Te}{\ensuremath{\T_{\envel}}}
\newcommand{\dotTe}{\ensuremath{\dot{\T}_{\envel}}}
\newcommand{\dotTR}{\ensuremath{\dot{\T}_{\return}}}
\newcommand{\nomPerc}{\rpipe}
\newcommand{\Ta}{\ensuremath{\T_{\amb}}}
\newcommand{\artTa}{\ensuremath{\hat{\T_{\amb}}}}
\newcommand{\avgTa}{\ensuremath{\overline{\T}_{\amb}}}
\newcommand{\avgTai}[1]{\ensuremath{\overline{\T}_{\amb,#1}}}
\newcommand{\avgTaCmp}{\avgTai{\compare}}
\newcommand{\avgTaExp}{\avgTai{\experiment}}
\newcommand{\Tsr}{\ensuremath{\T_{\sensor,\room}}}
\newcommand{\Ts}{\ensuremath{\T_{\sensor}}}
\newcommand{\dotTs}{\ensuremath{\dot{\T}_{\sensor}}}
\newcommand{\TRr}{\ensuremath{\T_{\return,\room}}}
\newcommand{\radius}{\ensuremath{r}}
\newcommand{\Isun}{\ensuremath{\Irr_{\SUN}}}
\newcommand{\Isundir}{\ensuremath{\Irr_{\SUN,\direct}}}
\newcommand{\forIsun}{\ensuremath{\tilde{\Irr}_{\SUN}}}
\newcommand{\forIsundir}{\ensuremath{\tilde{\Irr}_{\SUN,\direct}}}
\newcommand{\Qday}{\ensuremath{\Q_{\daily}}}
\newcommand{\efficHP}{\ensuremath{\eta_{\hp}}}
\newcommand{\copcarnot}{\ensuremath{\cop_\text{CARNOT}}}
\newcommand{\cophp}{\ensuremath{\cop_\hp}}
\newcommand{\fhp}{\ensuremath{f_\hp}}
\newcommand{\fgrid}{\ensuremath{f_\grid}}
\newcommand{\fcomf}{\ensuremath{f_\comf}}
\newcommand{\fcomfi}[1]{\ensuremath{f_{\comf,#1}}}
\newcommand{\zcomf}{\ensuremath{z_\comf}}
\newcommand{\fslack}{\ensuremath{f_\slack}}
\newcommand{\fcoTwo}{\ensuremath{f_{\coTwo}}}
\newcommand{\Pgplus}{\ensuremath{\Pg^+}}
\newcommand{\zhp}{\ensuremath{z_\hp}}
\newcommand{\zhpi}[1]{\ensuremath{z_{\hp,#1}}}
\newcommand{\Volt}{\mathcal{U}}
\newcommand{\Volta}{\Volt_\amb}
\newcommand{\forCloudCover}{\tilde{\perc}_\cloud}

\newcommand{\etahp}{\ensuremath{\eta_{\hp}}}
\newcommand{\Nwin}{\ensuremath{N}}

\newcommand{\TB}{\ensuremath{\T_{\brine}}}
\newcommand{\TF}{\ensuremath{\T_{\forward}}}
\newcommand{\nomTF}{\ensuremath{\hat{\T}_{\forward}}}
\newcommand{\TR}{\ensuremath{\T_{\return}}}
\newcommand{\Php}{\ensuremath{\pow_{\hp}}}
\newcommand{\Pg}{\ensuremath{\pow_{\grid}}}
\newcommand{\Ppv}{\ensuremath{\pow_{\pv}}}
\newcommand{\forPpv}{\ensuremath{\tilde{\pow}_{\pv}}}
\newcommand{\Pb}{\ensuremath{\pow_{\bess}}}
\newcommand{\Eb}{\ensuremath{\energy_{\bess}}}
\newcommand{\Pbc}{\ensuremath{\pow_{\bess, c}}}
\newcommand{\Pbd}{\ensuremath{\pow_{\bess, d}}}
\newcommand{\Papp}{\ensuremath{\pow_{\app}}}
\newcommand{\cg}{\price_{\grid}}
\newcommand{\dotQhp}{\ensuremath{\dotQ_{\hp}}}
\newcommand{\dotQsun}{\ensuremath{\dotQ_{\sun}}}
\newcommand{\Qhp}{\ensuremath{\Q_{\hp}}}
\newcommand{\dotQfh}{\ensuremath{\dotQ_{\fh}}}
\newcommand{\dotQsc}{\ensuremath{\hat{\dotQ}}}
\newcommand{\flowsc}{\ensuremath{\hat{\flow}}}
\newcommand{\dotQfhsc}{\ensuremath{\hat{\dotQ}_{\fh}}}
\newcommand{\Qfh}{\ensuremath{\Q_{\fh}}}
\newcommand{\heatFHRel}{\ensuremath{\dotQ_{\fh,\relative}}}
\newcommand{\QFHRel}{\ensuremath{\Q_{\fh,\relative}}}
\newcommand{\QFHRelj}[1]{\ensuremath{\Q_{\fh,\relative,#1}}}
\newcommand{\percj}[1]{\ensuremath{\perc_{#1}}}
\newcommand{\percFHRelj}[1]{\ensuremath{\perc_{\fh,\relative,#1}}}

\newcommand{\Ua}{\ensuremath{\U_{\area}}}
\newcommand{\Uf}{\ensuremath{\U_{\floor}}}
\newcommand{\Ur}{\ensuremath{\U_{\room}}}
\newcommand{\UR}{\ensuremath{\U_{\return}}}
\newcommand{\Uamb}{\ensuremath{\U_{\amb}}}
\newcommand{\Uw}{\ensuremath{\U_{\water}}}
\newcommand{\Ue}{\ensuremath{\U_{\envel}}}
\newcommand{\propU}{\hat{\U}}
\newcommand{\propUf}{\ensuremath{\propU_{\floor}}}
\newcommand{\propUr}{\ensuremath{\propU_{\room}}}
\newcommand{\propUR}{\ensuremath{\propU_{\return}}}
\newcommand{\propUa}{\ensuremath{\propU_{\amb}}}
\newcommand{\propUw}{\ensuremath{\propU_{\water}}}
\newcommand{\propUe}{\ensuremath{\propU_{\envel}}}
\newcommand{\URr}{\ensuremath{\U_{\return,\room}}}
\newcommand{\Usr}{\ensuremath{\U_{\sensor,\room}}}

\newcommand{\RA}{\ensuremath{\res_{\area}}}
\newcommand{\Rf}{\ensuremath{\res_{\floor}}}
\newcommand{\Ra}{\ensuremath{\res_{\amb}}}
\newcommand{\Rw}{\ensuremath{\res_{\water}}}
\newcommand{\Rsr}{\ensuremath{\res_{\sensor,\room}}}
\newcommand{\RR}{\ensuremath{\res_{\return}}}
\newcommand{\RRr}{\ensuremath{\res_{\return,\room}}}

\newcommand{\uarea}{\ensuremath{\su_{\area}}}
\newcommand{\uf}{\ensuremath{\spu_{\floor}}}
\newcommand{\ur}{\ensuremath{\spu_{\room}}}
\newcommand{\ua}{\ensuremath{\spu_{\amb}}}
\newcommand{\uw}{\ensuremath{\spu_{\water}}}
\newcommand{\ue}{\ensuremath{\spu_{\envel}}}
\newcommand{\utr}{\ensuremath{\spu_{\transport}}}
\newcommand{\uR}{\ensuremath{\spu_{\return}}}
\newcommand{\uRr}{\ensuremath{\spu_{\return,\room}}}

\newcommand{\Cr}{\ensuremath{\Capa_{\room}}}
\newcommand{\propCr}{\ensuremath{\hat{\Capa}_{\room}}}
\newcommand{\CR}{\ensuremath{\Capa_{\return}}}
\newcommand{\Cf}{\ensuremath{\Capa_{\floor}}}
\newcommand{\propCf}{\ensuremath{\hat{\Capa}_{\floor}}}
\newcommand{\Cw}{\ensuremath{\Capa_{\water}}}
\newcommand{\Ctr}{\ensuremath{\Capa_{\transport}}}
\newcommand{\Cs}{\ensuremath{\Capa_{\sensor}}}
\newcommand{\Ce}{\ensuremath{\Capa_{\envel}}}

\newcommand{\cw}{\ensuremath{c_{\water}}}
\newcommand{\cair}{\ensuremath{c_{\air}}}

\newcommand{\rhow}{\ensuremath{\rho_{\water}}}
\newcommand{\rpipe}{\ensuremath{\radius_{\pipe}}}

\newcommand{\dotQsrc}{\ensuremath{\dot{\Q}_{\source}}}
\newcommand{\Psrc}{\ensuremath{\pow_{\source}}}

\newcommand{\lenpipe}{\ensuremath{\len_{\pipe}}}
\newcommand{\lenroom}{\ensuremath{\len_{\room}}}
\newcommand{\Apipe}{\ensuremath{\A_{\pipe}}}

\newcommand{\Phpi}[1]{\ensuremath{\pow_{\hp,#1}}}


\newcommand{\Tri}[1]{\ensuremath{\T_{\room,#1}}}
\newcommand{\dotTri}[1]{\ensuremath{\dot{\T}_{\room,#1}}}

\newcommand{\Tfi}[1]{\ensuremath{\T_{\floor,#1}}}
\newcommand{\dotTfi}[1]{\ensuremath{\dot{\T}_{\floor,#1}}}

\newcommand{\Twi}[1]{\ensuremath{\T_{\water,#1}}}
\newcommand{\dotTwi}[1]{\ensuremath{\dot{\T}_{\water,#1}}}

\newcommand{\Tei}[1]{\ensuremath{\T_{\envel,#1}}}
\newcommand{\dotTei}[1]{\ensuremath{\dot{\T}_{\envel,#1}}}

\newcommand{\Ttri}[1]{\ensuremath{\T_{\transport,#1}}}
\newcommand{\dotTtri}[1]{\ensuremath{\dot{\T}_{\transport,#1}}}

\newcommand{\TRi}[1]{\ensuremath{\T_{\return,#1}}}
\newcommand{\dotTtRi}[1]{\ensuremath{\dot{\T}_{\return,#1}}}

\newcommand{\TrRefi}[1]{\ensuremath{\T_{\room,\reference,#1}}}

\newcommand{\Tai}[1]{\ensuremath{\T_{\amb,#1}}}
\newcommand{\TRj}[1]{\ensuremath{\T_{\return,#1}}}
\newcommand{\UAi}[1]{\ensuremath{\U_{\area,#1}}}
\newcommand{\Cri}[1]{\ensuremath{\Capa_{\room,#1}}}
\newcommand{\Cfi}[1]{\ensuremath{\Capa_{\floor,#1}}}
\newcommand{\Cwi}[1]{\ensuremath{\Capa_{\water,#1}}}
\newcommand{\Ctri}[1]{\ensuremath{\Capa_{\transport,#1}}}
\newcommand{\Ti}[2]{\ensuremath{\T_{#1,#2}}}
\newcommand{\Ci}[2]{\ensuremath{\Capa_{#1,#2}}}
\newcommand{\Ui}[2]{\ensuremath{\U_{#1,#2}}}
\newcommand{\dotQij}[2]{\ensuremath{\dotQ_{#1,#2}}}
\newcommand{\dotUi}[2]{\ensuremath{\dotU_{#1,#2}}}
\newcommand{\dotTi}[2]{\ensuremath{\dotT_{#1,#2}}}
\newcommand{\Rwi}[1]{\ensuremath{\res_{\water,#1}}}
\newcommand{\cgi}[1]{\ensuremath{\price_{\grid,#1}}}
\newcommand{\ci}[1]{\ensuremath{c_{#1}}}
\newcommand{\lenpj}[1]{\ensuremath{\len_{\room,#1}}}
\newcommand{\Apj}[1]{\ensuremath{\Area_{\pipe,#1}}}
\newcommand{\Arj}[1]{\ensuremath{\Area_{\room,#1}}}
\newcommand{\dtrj}[1]{\ensuremath{\dt_{\room,#1}}}
\newcommand{\dttrj}[1]{\ensuremath{\dt_{\transport,#1}}}
\newcommand{\dotQfhj}[1]{\ensuremath{\dotQ_{\fh,#1}}}
\newcommand{\nomvali}[1]{\ensuremath{\nomqprop_{#1}\val_{#1}}}
\newcommand{\dotQhpi}[1]{\ensuremath{\dotQ_{\hp,#1}}}

\newcommand{\Ufi}[1]{\ensuremath{\U_{\floor,#1}}}
\newcommand{\Uai}[1]{\ensuremath{\U_{\amb,#1}}}
\newcommand{\Uri}[1]{\ensuremath{\U_{\room,#1}}}
\newcommand{\Uwi}[1]{\ensuremath{\U_{\water,#1}}}
\newcommand{\Uei}[1]{\ensuremath{\U_{\envel,#1}}}
\newcommand{\RAi}[1]{\ensuremath{\res_{\area,#1}}}
\newcommand{\Rfi}[1]{\ensuremath{\res_{\floor,#1}}}
\newcommand{\Rai}[1]{\ensuremath{\res_{\amb,#1}}}
\newcommand{\Rri}[1]{\ensuremath{\res_{\room,#1}}}
\newcommand{\Rtri}[1]{\ensuremath{\res_{\transport,#1}}}
\newcommand{\flowi}[1]{\ensuremath{q_{#1}}}
\newcommand{\Trefi}[1]{\ensuremath{\T_{\room,\reference,#1}}}
\newcommand{\QFHReli}[1]{\ensuremath{\Q_{\fh,\relative,#1}}}
\newcommand{\dotQFHReli}[1]{\ensuremath{\dotQ_{\fh,\relative,#1}}}

\newcommand{\trp}{\ensuremath{\text{T}}}

\newcommand{\cgex}{\cgi{\expo}}
\newcommand{\cgim}{\cgi{\impo}}

\newcommand{\Hp}{\ensuremath{h_p}}

\newcommand{\fhpmin}{\ensuremath{\underline{f_{\hp}}}}
\newcommand{\fhpmax}{\ensuremath{\overline{f_{\hp}}}}
\newcommand{\Phpmin}{\ensuremath{\underline{\Php}}}
\newcommand{\Phpmax}{\ensuremath{\overline{\Php}}}
\newcommand{\fhpi}[1]{\ensuremath{f_{\hp,#1}}}

\newcommand{\Am}{\ensuremath{\mathbf{A}}}
\newcommand{\Bm}{\ensuremath{\mathbf{B}}}
\newcommand{\Cm}{\ensuremath{\mathbf{C}}}
\newcommand{\Dm}{\ensuremath{\mathbf{D}}}
\newcommand{\Em}{\ensuremath{\mathbf{E}}}
\newcommand{\Nm}{\ensuremath{\mathbf{N}}}
\newcommand{\Um}{\ensuremath{\mathbf{U}}}
\newcommand{\Vm}{\ensuremath{\mathbf{V}}}
\newcommand{\ident}{\ensuremath{\mathbf{I}}}
\newcommand{\Wm}{\ensuremath{\mathbf{W}}}
\newcommand{\Zm}{\ensuremath{\mathbf{Z}}}
\newcommand{\Pm}{\ensuremath{\mathbf{P}}}
\newcommand{\Tm}{\ensuremath{\mathbf{T}}}
\newcommand{\Qm}{\ensuremath{\mathbf{Q}}}
\newcommand{\Rm}{\ensuremath{\mathbf{R}}}

\newcommand{\Acal}{\ensuremath{\mathcal{A}}}
\newcommand{\Bcal}{\ensuremath{\mathcal{B}}}
\newcommand{\Ccal}{\ensuremath{\mathcal{C}}}
\newcommand{\Dcal}{\ensuremath{\mathcal{D}}}
\newcommand{\Ecal}{\ensuremath{\mathcal{E}}}
\newcommand{\Ncal}{\ensuremath{\mathcal{N}}}
\newcommand{\Ucal}{\ensuremath{\mathcal{U}}}
\newcommand{\Vcal}{\ensuremath{\mathcal{V}}}
\newcommand{\Wcal}{\ensuremath{\mathcal{W}}}
\newcommand{\Zcal}{\ensuremath{\mathcal{Z}}}
\newcommand{\Pcal}{\ensuremath{\mathcal{P}}}
\newcommand{\Tcal}{\ensuremath{\mathcal{T}}}

\newcommand{\avm}{\ensuremath{\mathbf{a}}}
\newcommand{\Sigm}{\ensuremath{\mathbf{\Sigma}}}
\newcommand{\Aouter}{\ensuremath{\A\A^*}}
\newcommand{\Ainner}{\ensuremath{\A^*\A}}
\newcommand{\ARinner}{\ensuremath{\A^\trp\A}}
\newcommand{\ZRinner}{\ensuremath{\Z^\trp\Z}}
\newcommand{\uv}{\ensuremath{\mathbf{u}}}
\newcommand{\vv}{\ensuremath{\mathbf{v}}}
\newcommand{\qv}{\ensuremath{\mathbf{q}}}
\newcommand{\xv}{\ensuremath{\mathbf{x}}}
\newcommand{\yv}{\ensuremath{\mathbf{y}}}
\newcommand{\zv}{\ensuremath{\mathbf{z}}}
\newcommand{\nv}{\ensuremath{\mathbf{n}}}
\newcommand{\bv}{\ensuremath{\mathbf{b}}}
\newcommand{\cvec}{\ensuremath{\mathbf{c}}}
\newcommand{\dv}{\ensuremath{\mathbf{d}}}

\newcommand{\rv}{\ensuremath{\mathbf{r}}}
\newcommand{\alphav}{\ensuremath{\mathbf{\alpha}}}

\newcommand{\realv}[1]{\ensuremath{\mathbb{R}^{#1}}}
\newcommand{\binv}[1]{\ensuremath{\mathbb{B}^{#1}}}
\newcommand{\iter}{\ensuremath{j}}

\newcommand{\placehold}{\ensuremath{\{\cdot\}}}

\newcommand{\rhc}{relative heating coefficient}
\newcommand{\RHC}{RHC}
\newcommand{\rmse}{root mean square error}
\newcommand{\RMSE}{RMSE}
\newcommand{\rhcs}{\beta}
\newcommand{\rhcsj}[1]{\ensuremath{\beta_{#1}}}
\newcommand{\rhcsq}{\rhcs_{\dotm}}
\newcommand{\rhcsqj}[1]{\rhcs_{\dotm,#1}}
\newcommand{\rhcsA}{\rhcs_{\Ar}}
\newcommand{\rhcsAj}[1]{\rhcs_{\Ar,#1}}
\newcommand{\rhcsAqj}[1]{\rhcs_{\Ar\dotm,#1}}

\newcommand{\dtsim}{\ensuremath{dt}}
\newcommand{\percPpart}{\ensuremath{\alpha_\pipe}}

\newcommand{\gParam}{\ensuremath{g}}
\newcommand{\fParam}{\ensuremath{\gParam_\floor}}
\newcommand{\eParam}{\ensuremath{\gParam_\envel}}
\newcommand{\rParam}{\ensuremath{\gParam_\room}}
\newcommand{\pParam}{\ensuremath{\gParam_\pipe}}
\newcommand{\qParam}{\ensuremath{\gParam_\dotm}}
\newcommand{\sunParam}[1]{\ensuremath{\gParam_{\SUN,#1}}}
\newcommand{\propsunParam}[1]{\ensuremath{\hat{\gParam}_{\SUN,#1}}}

\newcommand{\eqDist}{0.5cm}
\newcommand{\numPart}{\ensuremath{N_\partition}}
\newcommand{\numTransportPart}{\ensuremath{M_\partition}}
\newcommand{\Ncirc}{N}
\newcommand{\numRooms}{\ensuremath{\Ncirc_r}}
\newcommand{\Nhor}{\ensuremath{N}}
\newcommand{\Nday}{\ensuremath{N_{\daily}}}
\newcommand{\Ncmp}{N_{\compare}}
\newcommand{\share}{\mathcal{P}}
\newcommand{\shareFHRelj}[1]{\ensuremath{\share_{\fh,\relative,#1}}}
\newcommand{\maxLike}{Maximum Likelihood}
\newcommand{\ML}{ML}
\newcommand{\leastSq}{Least Squares}
\newcommand{\LS}{LS}
\newcommand{\UKF}{UKF}

\newcommand{\mpc}{Model Predictive Control}
\newcommand{\MPC}{MPC}

\newcommand{\mimpc}{Mixed Integer Model Predictive Control}
\newcommand{\MIMPC}{MIMPC}

\newcommand{\milp}{Mixed Integer Linear Programming}
\newcommand{\MILP}{MILP}

\newcommand{\miocp}{Mixed Integer Optimal Control Problem}
\newcommand{\MIOCP}{MIOCP}

\newcommand{\ocp}{optimal control problem}
\newcommand{\Ocp}{Optimal control problem}
\newcommand{\OCP}{OCP}

\newcommand{\efficiencyModel}{efficiency model}

\newcommand{\zoh}{Zero Order Hold}
\newcommand{\ZOH}{ZOH}

\newcommand{\tes}{thermal energy storage}
\newcommand{\TES}{TES}

\newcommand{\ebat}{electric battery}
\newcommand{\BESS}{BAT}

\newcommand{\tra}{\ensuremath{\text{T}}}

\newcommand{\nzeb}{near zero-emission building}
\newcommand{\NZEB}{NZEB}
\newcommand{\backend}{backend}

\newcommand{\atow}{air-to-water}

\newcommand{\lkf}{Linear Kalman Filter}
\newcommand{\LKF}{LKF}

\newcommand{\singleFam}{single-family}
\newcommand{\SingleFam}{Single-family}
\newcommand{\dataSet}{data-set}
\newcommand{\DataSet}{Data-set}
\newcommand{\spaceHeating}{space-heating}
\newcommand{\SpaceHeating}{Space-heating}
\newcommand{\edmdc}{Extended Dynamic Mode Decomposition with Control}
\newcommand{\EDMDC}{EDMDc}

\newcommand{\DHW}{DHW}
\newcommand{\dhw}{domestic hot water}

\newcommand{\eu}{European Union}
\newcommand{\EU}{EU}

\newcommand{\mld}{Mixed Logic Dynamics}
\newcommand{\MLD}{MLD}

\newcommand{\OP}{optimization problem}
\newcommand{\op}{\ensuremath{\text{OP}}}

\newcommand{\nordpool}{Nord Pool}

\newcommand{\currency}{\euro}
\newcommand{\cookingPeak}{evening peak}
\newcommand{\CookingPeak}{Evening peak}

\newcommand{\loadShift}{load shift}
\newcommand{\LoadShift}{Load shift}

\newcommand{\loadshifting}{\loadShift ing}
\newcommand{\Loadshifting}{\LoadShift ing}

\newcommand{\dsm}{demand side management}
\newcommand{\Dsm}{Demand side management}
\newcommand{\DSM}{DSM}

\newcommand{\smartGrid}{smart grid}
\newcommand{\SmartGrid}{Smart grid}

\newcommand{\savingRateText}{saving rate}
\newcommand{\SavingRateText}{Saving rate}

\newcommand{\bilinear}{bilinear}

\newcommand{\ipcc}{The Intergovernmental Panel on Climate Change}
\newcommand{\IPCC}{IPCC}

\newcommand{\tsos}{Transport Service Operator}
\newcommand{\TSOs}{TSO}
\newcommand{\dsos}{Distribution Service Operator}
\newcommand{\DSOs}{DSO}

\newcommand{\daySetCmp}{\mathcal{D}_\compare}
\newcommand{\daySetExp}{\mathcal{D}_\experiment}
\newcommand{\SSdx}[1]{\xv_{#1+1} = \Am \xv_{#1} + \Bm \uv_{#1} + \Em \dv_{#1}}
\newcommand{\SScx}[1]{\dot{\xv}_{#1} = \Am \xv_{#1} + \Bm \uv_{#1} + \Em \dv_{#1}}

\newcommand{\iterOp}{i}

\newcommand{\figSpace}{-7mm}
\newcommand{\SecSpace}{-3mm}
\newcommand{\tableSpace}{-6mm}

\newcommand{\degC}{\si{\degreeCelsius}}
\newcommand{\kwh}{\si{\kilo\watt\hour}}

\newcommand{\figUpper}{\text{Upper: }}
\newcommand{\figLower}{\text{Lower: }}
\newcommand{\figLeft}{\text{Left: }}
\newcommand{\figRight}{\text{Right: }}
\newcommand{\figLowerLeft}{\text{Lower left: }}
\newcommand{\figLowerRight}{\text{Lower right: }}
\newcommand{\figUpperLeft}{\text{Upper left: }}
\newcommand{\figUpperRight}{\text{Upper right: }}

\newcommand{\questionTwo}{\textit{What impact does the controller have on indoor climate?}}
\newcommand{\questionThree}{\textit{Is it possible to load the floor with heat during high irradiation periods from a low hanging sun without severely compromising comfort?}}
\newcommand{\questionFour}{\textit{How well does the efficiency model of the heat pump predict the actual efficiency?}}
\newcommand{\questionFive}{\textit{How well does the efficiency model of the heat pump predict the actual efficiency?}}

\newcommand{\neogrid}{Neogrid Technologies}

\newcommand{\comfLvlTriangleOne}{{\color{ref1}\FilledBigTriangleUp}}
\newcommand{\comfLvlTriangleTwo}{{\color{ref2}\FilledBigTriangleUp}}
\newcommand{\comfLvlTriangleThree}{{\color{ref3}\FilledBigTriangleUp}}
\newcommand{\comfLvlTriangleFour}{{\color{ref4}\FilledBigTriangleUp}}

\newcommand{\rsquaredInit}{0.92}
\newcommand{\rsquaredUpd}{0.96}

\newcommand{\percSaving}{12}
\newcommand{\numTestDays}{97}
\newcommand{\totalSavings}{28.4}
\newcommand{\savingRate}{0.29}
\newcommand{\yearsEarning}{34}
\newcommand{\comfTOnePercSave}{32.8}
\newcommand{\comfTwoPercSave}{29.1}
\newcommand{\comfThreePercSave}{2.3}
\newcommand{\comfFourPercSave}{17.4}
\newcommand{\extraPeakCost}{34.9}
\newcommand{\extraPeakElec}{60.0}
\newcommand{\avgEveningPrice}{0.27}
\newcommand{\copPeak}{4.2}
\newcommand{\costAfter}{16.1}
\newcommand{\savePeakBlock}{18.8}
\newcommand{\percPeakSave}{66}
\newcommand{\comfOneCostPerc}{32.8}
\newcommand{\comfTwoCostperc}{29.1}
\newcommand{\comfThreeCostPerc}{2.3}
\newcommand{\comfFourCostPerc}{17.4}
\newcommand{\comfOneCostPercNoDec}{32}
\newcommand{\comfTwoCostpercNoDec}{29}
\newcommand{\comfThreeCostPercNoDec}{2}
\newcommand{\comfFourCostPercNoDec}{17}
\newcommand{\expEndDate}{2023-03-06}
\newcommand{\comfFourNumDays}{32}
\newcommand{\comfOnePercSaveNoDec}{33}
\newcommand{\comfTwoPercSaveNoDec}{29}
\newcommand{\comfThreePercSaveNoDec}{2}
\newcommand{\comfFourPercSaveNoDec}{17}

%% file: sections/introduction/introduction.tex
\section{Introduction}
\begin{figure}[h]
	\centering
	\includegraphics[width=0.925\columnwidth]{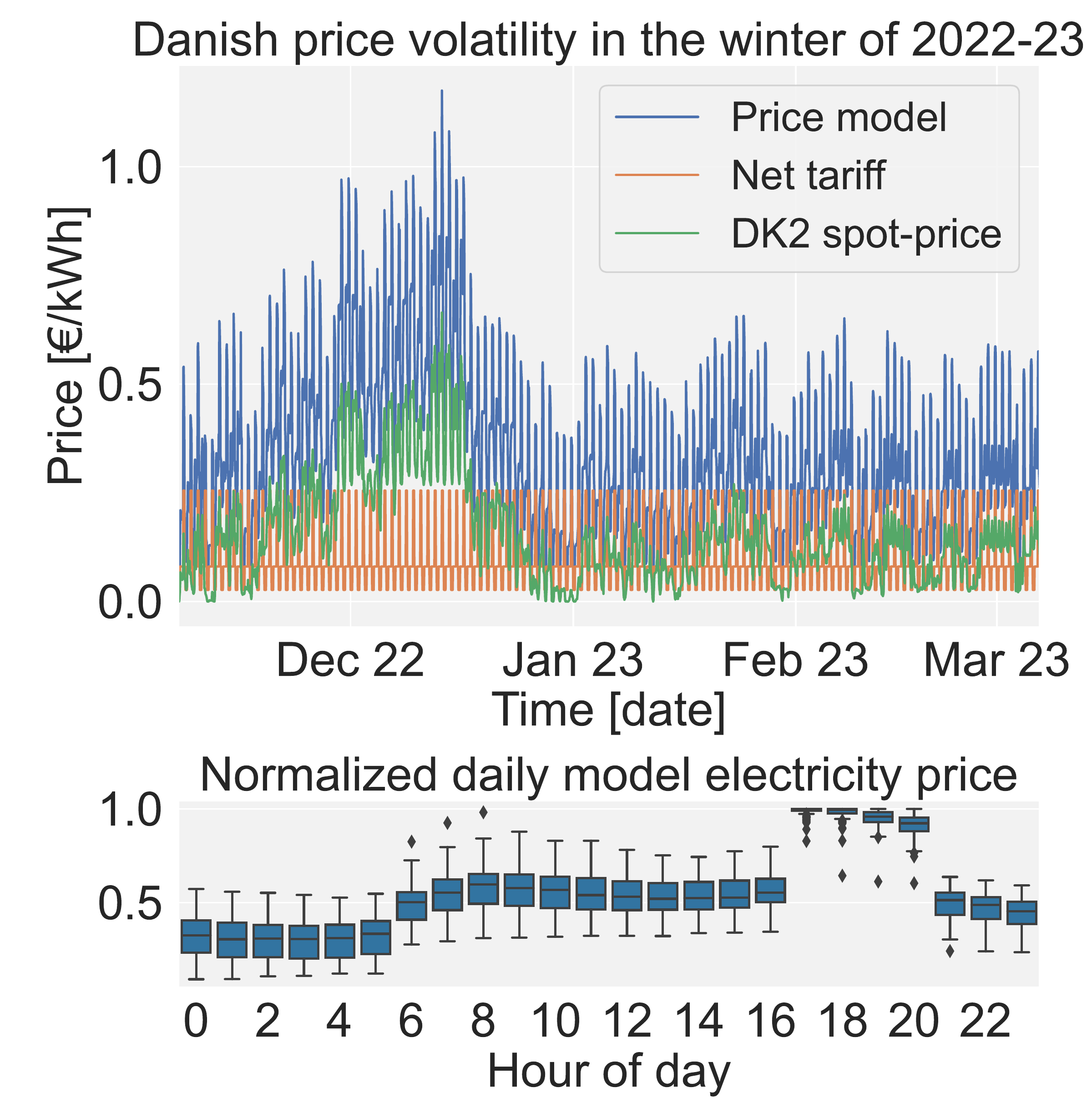}
	\label{fig_price}
	\caption{\figUpper The Nord Pool market spot price plotted together with the Danish tariff Model 3.0. The price model is used for the economical evaluation of the experiment. \figLower The normalized ($x_i/\max(x)$) daily price development of the price model.}
\end{figure}

A fast and determined transition to a carbon neutral economy is more urgent than ever.
The summary for policy makers associated with the $6^{th}$ annual report from \ipcc\ reads: \textit{"All global modelled pathways that limit warming to $1.5^{\circ}C (>50\%)$ with no or limited overshoot, and those that limit warming to $2^\circ C (>67\%)$ involve rapid and deep and in most cases immediate [Green house gas]}  \textit{emission reductions in all sectors"} \cite{ipcc6_wg3}. This means that not only long term solutions, but also existing solutions need to be implemented, immediately. Space heating is major energy consumer with potential for large reductions both short and long term. 
The focus here is on \singleFam\ houses, since they pose a particular grand challenge for the overall savings potential in the space heating sector. \SingleFam\ houses are small but many in numbers, meaning that they make up a large share of the sector. Estimates indicate that about 55\% of Danish heated area belongs to \singleFam\ houses \cite{danish_energy_agency_data_nodate}. Further complicating the issue is that a majority of the \singleFam\ houses are owned by the residents themselves \citenum{statistics_denmark_boligbestanden_2021}. This is not bad in itself---self-ownership has many socioeconomic benefits---but it does mean that any solution introduced to a \singleFam\ house has to be highly cost-beneficial in order to get the individual owners to invest in energy upgrades. A popular investment, seen across the \eu, is to acquire a heat pump (\hp). In the period 2005 to 2020, sales increased from about 0.5 mill. to 1.62 mill units sold with air-sourced being the most popular type \cite{heat_pump_market}. The rise in heat pumps is only one factor in an increasingly electrified economy, which starts to put strain on the electric grid with peak loads threatening stability and capacity. 
In Denmark the response is a new network tariff model for electricity, \textit{tarifmodel 3.0}, which was introduced on the $1^{\text{st}}$ of January, 2023 \cite{tarifmodel3}. This model allows the DSO(grid)-operators to differentiate the end-user tariffs substantially over the course of the day in order to nudge the end-user into changing their consumption away from peak load periods and increase demand at night. This situation also impacts households heated by an electric heat pump who, although having some tax-benefits, still have to pay the full grid-tariffs. In other words, the owners need to change their heating habits or face the cost of heating in expensive periods. Many danish households are already on a time-varying price which is based on the \nordpool\ hourly spot-market and time-of-use distribution prices. Adding the two price schemes together means that the difference between high and low prices within a day can be several times larger than the lower price, as seen in \figref{fig_price}. This can create some very costly situations, but also opportunities for cost savings. 
\vspace{-1mm}
One such opportunity is to utilize a price-aware controller to load shift by boosting heat production (charge) in low cost periods and decrease (discharge) in high cost periods either with help of an energy storage \cite{kelly_performance_2014, bechtel_influence_2020} or directly using the thermal mass of the building itself \cite{amato_dual-zone_2023, sanchez_ramos_potential_2019, hu_price-responsive_2019}. A variable-speed heat pump can be used to boost heat by increasing the compressor speed, but this comes with a significant loss of efficiency (\textit{coefficient of performance}, \cop). Further, the \cop\ of an \atow\ \hp\ is highly dependent on ambient temperature, which means that not only price but also weather factors need to be considered as well.

A method suitable for automating heat \loadshifting\ is \mpc\ (\MPC) \cite{amato_dual-zone_2023}. During the last 30 years, \MPC\ has been studied extensively in the context of building control due to its structural ability to integrate building dynamics, heating system and environmental aspects into an \ocp\ (\OCP) formulation capable of handling both constraints and discrete states. A range of different versions of \MPC\ have been suggested: Deterministic \MPC, Stochastic \MPC, Robust \MPC, Learning \MPC, Offset-free \MPC, Implicit \MPC\ and Explicit \MPC\ \cite{drgona_all_2020}. While the studies are numerous, the method has so far failed to make a broad impact on the space heating sector. The reported reasons are installation costs of sensor and actuators, model development costs \cite{sturzenegger_model_2016} and user-acceptance.

Although the number of long term (beyond 30 days) building scale demonstrations are few compared to simulation studies, a selection of noteworthy examples do exist. In \cite{sturzenegger_model_2016} 
a 6000 \si{\meter\squared}, occupied office building in Switzerland both in periods during winter and summer which combined into about 30 weeks of testing. While reporting that the control itself was a success, the author questions whether the method is mature enough to be implemented in similar buildings. In \cite{de_coninck_practical_2016} two heat pumps and a gas boiler were controlled in a 960 \si{\meter\squared} occupied office building in Brussels during the winter of 2014-2015 reporting cost savings of 30\% while improving comfort. In Halifax, Canada, a 10000 \si{\meter\squared} university building was controlled using \MPC\ for four months with reported savings of 29\% electricity and 63\% heat \cite{hilliard_experimental_2017}. In the category \singleFam\ houses, \cite{pedersen_central_2013} controlled four houses for 5 months and reported an average cost reduction of 9\% when compared to 7 benchmark houses and in \cite{muller_large-scale_2019} \hp s in 300 homes were ON/OFF throttled to reduce peak loads.  The low number of residential experiments is likely due to the low potential for savings, which disqualifies large implementation costs. The requirement for simple solutions have spurred a branch of low-complexity\footnote{Low-complexity is meant as relative to solutions where the heat consumption of each heating zone is known.} \MPC\ e.g. with only one central heat meter as in \cite{amato_dual-zone_2023}. Recent studies \cite{amato_dual-zone_2023, vogler-finck_inverse_2019} have demonstrated the basic feasibility of such schemes, but both studies point out that longer evaluation periods are needed to reliably verify their practical usefulness. Furthermore, occupancy in \singleFam\ houses is a fundamentally different condition from office buildings due to the invasive nature of sensor feedback on the occupants´ behavior, which must also be addressed.

Our contribution in this paper is a \textit{\numTestDays\ day long} study demonstrating a price responsive, low-complexity, hierarchical Mixed-Integer \MPC\ control scheme on an \textit{occupied} \singleFam\ house featuring an \atow\ heat pump and floor heating (\fh). The controller is designed to minimize costs by shifting heating loads according to the electricity price signal together with other predictable and/or measurable factors. The controller is developed as a comparably low-complexity solution which only makes use of an internet connected control unit, a central heat meter, electricity meters, and room thermostats. Further, the weather forecast is provided by a weather service and the model is a single zone model which is based on a weighted average room temperature for the entire house. The controller is deliberately designed not to make use of explicit occupancy information, in order to protect the occupants´ right to privacy. The main findings from the experiment are: the near zero emission house demonstrated a high level of flexibility with respect to time-of-heating. Further, it is possible to boost the floors with heat during intensive sun radiation periods (when there is plenty of own-produced PV electricity) without further deteriorating the comfort. Controlling the upper layer using an area weighted average building temperature has shown to be unproblematic with respect to comfort in the test house. 

The layout for the rest of the paper is as follows. Section \ref{sec_system} presents the case, an overview of the heating side and the electrical side viewed from a control perspective. Section \ref{sec_control_strategy} presents the hierarchical control strategy, starting with the supervisory controller and followed by the mid-level controllers. Section \ref{sec_model} contains the models used in the paper. Section \ref{sec_experiment} describes the experiment before the results are presented in Section \ref{sec_results}. As the results are based on real data, Section \ref{sec_result_interpret} is dedicated to the authors' interpretation of the results. Finally, a common discussion section followed by conclusion in Sections \ref{sec_discussion} and \ref{sec_conclusion}, respectively.

\input{sections/introduction/test_house}

%% file: sections/method/method.tex
\section{System}
\label{sec_system}
\input{sections/method/system}
\input{sections/method/control_strategy}
\input{sections/method/mpc}
\input{sections/method/valve_control}
\input{sections/method/heat_pump_control}
\input{sections/method/model}
\input{sections/method/house_model}
\input{sections/method/state_est}
\input{sections/method/heat_pump_model}
\input{sections/method/pv_model}
\input{sections/method/price}
\input{sections/method/validation}


%% file: sections/method/system.tex
This section starts with an introduction to the case followed by an overview of the heating system and electrics before delving into the control retro-fit. The relevant signals are listed in Table~\ref{table_relevant_signals}.
\subsection{Case study}
\label{sec_test_house}
\begin{figure}[h]
	\centering
	\includegraphics[width=.75\columnwidth]{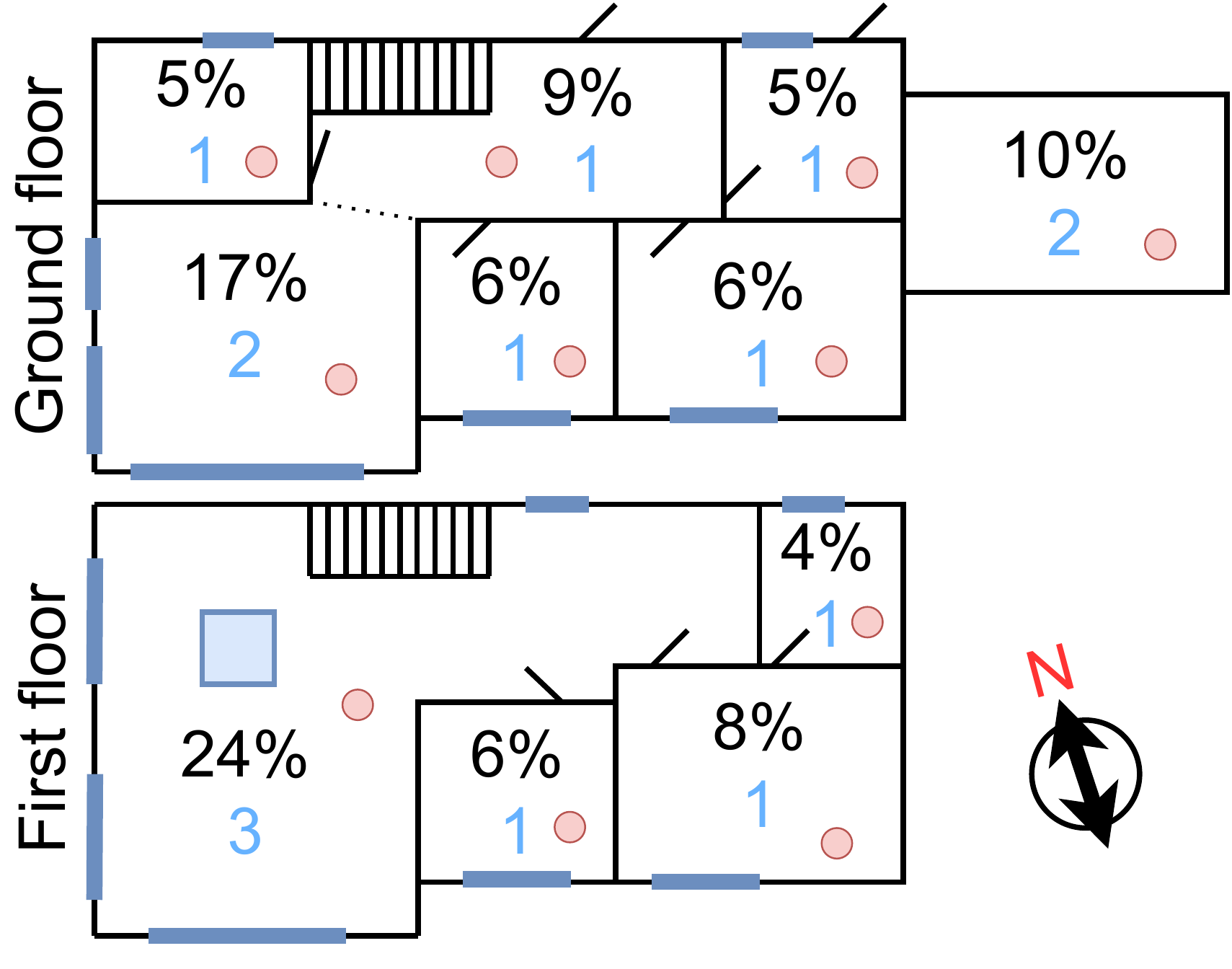}
	\label{fig_blueprint}
	\caption{Blueprint of the floor plan with area distribution. Blue numbers indicate the number of \fh\ circuits in each zone. Red circles show the presence of a room thermostat.}
\end{figure}

The case is a \SI{230}{\meter\squared}, two-story \singleFam\ house from 2018; see \figref{fig_blueprint}. According to the Danish building regulation, it is classified as a low energy class building (BR2020), which, among other requirements, implies a maximum annual heat demand of \SI{20}{\kilo\watt\hour\per\meter\squared} \cite{br18}. It is located on Sjælland (Zealand) in Denmark, with a south view over the sea. A south facing photovoltaic system is placed on the roof with a measured peak output of \SI{4}{\kilo\watt} in end of December and \SI{5.5}{\kilo\watt} in June. Space heating and domestic hot water is provided by a \textit{Bosch Compress 6000AW} (air-to-water) heat pump with a nominal capacity of \SI{7}{\kilo\watt}. Domestic hot water takes priority over space heating. Based on measured data the nominal electric consumption ranges from \SI{200}{\watt} to \SI{2500}{\watt}. Floor heating, embedded in concrete, is installed throughout the house. The floor heating system is controlled by a Wavin controller and consists of 15 circuits delivering heat to 11 heating zones. Each zone has one thermostat assigned, meaning that if more circuits are supplying the same zone all valves in the particular zone opens when heat is requested. The circuits are ON/OFF controlled based on deviations from the temperature reference provided for each zone. The heat pump is controlled by an ambient temperature compensated heat curve. The household has a variable electricity price contract, which is based on the \nordpool\ market spot-price.
\newcommand{\tablePercOne}{0.24}
\begin{table}[H]
\centering
\caption{Relevant signals}
\begin{tabular}{p{0.14\linewidth} p{0.1\columnwidth} 
p{0.57\columnwidth}}
\toprule[1.5pt]
Variable  & Unit & Description \\ 
\midrule[1.5pt]
 $\dotQhp$ & $[\si{\watt}]$ & Heat flow output from \hp \\
 $\Qhp$ & $[\kwh]$ & Heat output from \hp \\ 
 $\Php$ & $[\si{\watt}]$ & Electric power input to \hp \\
 $\Ehp$ & $[\kwh]$ & Electric energy input to \hp \\
 $\Ppv$ & $[\si{\watt}]$ & Electric power output from photovoltaic \\
 $\Pg$ & $[\si{\watt}]$ & Electric power from grid \\
 $\Papp$ & $[\si{\watt}]$ & Household electric power consumption \\
 $\TR$ & $[\si{\degreeCelsius}]$ & Return temperature to \hp \\
  $\TRi{i}$ & $[\si{\degreeCelsius}]$ & Return Temperature from \fh\ circuit $i$ \\
    $\Tri{j}$ & $[\si{\degreeCelsius}]$ & Air temperature in room $j$ \\
    $\Trefi{j}$ & $[\si{\degreeCelsius}]$ & Reference temp. in room $j$ \\
    $\Tfi{j}$ & $[\si{\degreeCelsius}]$ & Estimated floor temperature in room $j$ \\
    $\val_{i}$ & $[\si{\degreeCelsius}]$ & ON/OFF Valve setting for circuit $i$ \\
    $\flow$ & $[\si{\kilogram\per\second}]$ & Total flow into the \fh\ system \\
    $\Volta$ & $[\si{\volt}]$ & Output voltage from ambient temperature sensor \\
\bottomrule[1.5pt]
\end{tabular}
\label{table_relevant_signals}
\end{table}
\subsection{Heating system}
\begin{figure}[H]
	\centering
	\includegraphics[width=1\columnwidth]{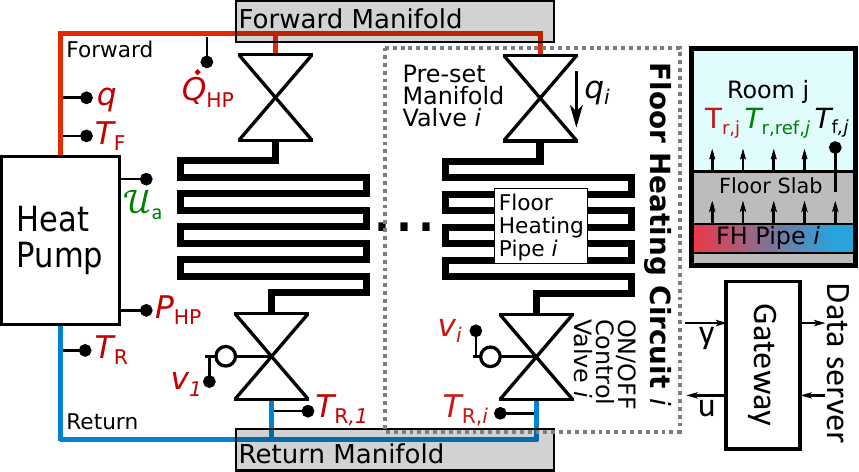}
	\caption{The diagram of the heating system in the house. The colors represent the kind of signal: red for measured variables, green for control inputs and black for estimated variables. Pipes on warm side are red and cold side blue.}
        \label{fig_system_diagram}
\end{figure}

\figref{fig_system_diagram} shows the heating system with associated signals. 
Note that heat flow to the floor heating system, \dotQhp, and electric power consumption of the heat pump, \Php, are measured.
The \hp\ feeds the floor heating system with water, which in turn deliver the heat to the heating zones.
\subsection{Electricity}
\begin{figure}[H]
	\centering
	\includegraphics[width=0.92\columnwidth]{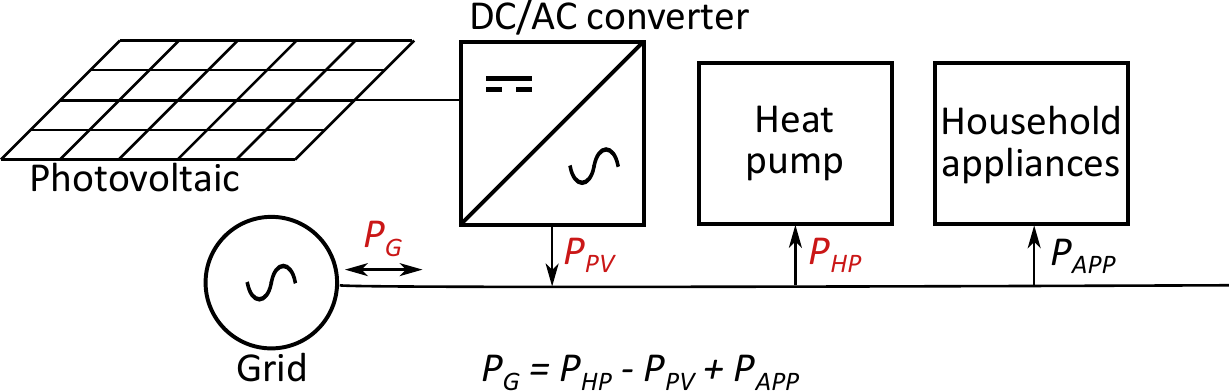}
	\caption{The internal electricity grid of the house expressed in power. Red variables are measured quantities.}
 	\label{fig_elec_diagram}
\end{figure}

The household electric grid is shown in Figure \ref{fig_elec_diagram}. The main units are photovoltaic panels and the heat pump which have separate electricity meters. The other household appliances are aggregated into an unknown disturbance.
\input{sections/method/hardware.tex}
\subsection{Object oriented description of commercial domestic heat pump}
\label{sec_hp_object}
In this section common properties for commercial \atow\ \hp's are listed together with references to how they are modelled in the literature.
\begin{enumerate}
\item \textbf{ON/OFF indicator}: the \hp\ turns off when heat demand is absent. In this work the indicator variable $\deltahp \in \{0,1\}$ is one when the \hp\ is ON and zero if OFF \cite{lee_mixed-integer_2019, mayer_management_2015}.
\item \textbf{Minimum load:} the minimum load and operation range of a variable speed \hp\ is often considered and modelled as a set $\Php \in \{0\}\cup [\Phpmin, \Phpmax]$ \cite{kuboth_economic_2019,lee_mixed-integer_2019, maier_assessing_2022}.
\item \textbf{Coefficient of performance:} the coefficient of performance (\cop) is the ratio between the produced heat and consumed input energy (here electricity). It is often modelled as a static function, $\fhp: \mathbb{R} \to \mathbb{R}$.
\item \textbf{Down-time:} to avoid start-up cycling some \hp s feature a (sometimes adaptive) down-time period measured in hours. To incorporate this a model for minimum up- and down-time can be included \cite{mayer_management_2015, lohr_mpc_2021}.
\item \textbf{Limit on rate off change:} the internal controllers of a domestic \hp\ sometimes prevent it from changing state too rapidly.
\item \textbf{Domestic hot water production:} the \hp\ switches between providing space heat and domestic hot water. Domestic hot water is often prioritized.
\item \textbf{Discrete compressor speed steps:} the compressor speed is often operated at certain steps rather than continuous action. Some speeds are excluded as resonance with the casing can cause noise.
\item \textbf{Low pass filter on ambient temperature signal:} it is common practice that commercial \hp s apply a low pass filter to the ambient temperature signal before it is provided to the internal controllers.
\item \textbf{Defrosting:} an \atow\ \hp\ needs to defrost the evaporator regularly in order to function properly. This event is treated as a random process which takes priority.
\end{enumerate}
It is desirable that any \MPC\ operating an \hp\ can handle the listed properties.

%% file: sections/method/hardware.tex
\subsection{Retrofit architecture}
\begin{figure}
	\centering
	\includegraphics[width=0.9\columnwidth]{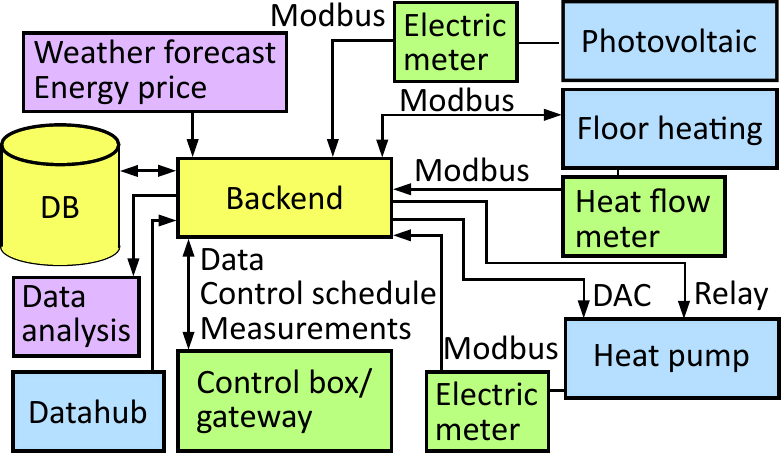}
	\caption{Shows the overview of the hardware and communication protocols. Blue color is for preinstalled hardware, green for installed sensors, yellow is the off-site infrastructure and purple are data services.}
 	\label{fig_hardware}
\end{figure}

The retrofit architecture, which is built and implemented by \neogrid, is seen in \figref{fig_hardware}. 
The infrastructure consists of an onsite part and a \backend\ with the control box acting as gateway between them. The \backend\ is responsible for refining, organizing, downloading data from weather and price services, and storing data, which is used for analysis and model fitting. The control box is responsible for providing control signals and collecting measurements from all units. In this case, it means to provide the artificial ambient temperature overwrite, via a digital to analog converter (DAC) and blocking the compressor using a relay. The datahub (Eloverblik) is an online platform, provided by the danish publicly owned company Energinet, where electricity-customers can get an overview of their consumption or share data with third-party. We use the BACnet and Modbus protocols to communicate with the floor heating controller for collecting room temperatures and other data from the floor heating system, and sending set-points to the valves.

%% file: sections/method/control_strategy.tex
\section{Control}
\label{sec_control_strategy}
The control objective is to provide the required comfort level at the lowest cost feasible. To accomplish this, the controller needs to make two high-level control actions. First, it must choose the heat pump heat flow $\dotQhp(t) \in \mathbb{R}$ and the \fh\ water flow $q(t) \in \mathbb{R}$. Second, it must guide the water to the most suitable rooms. It is not possible to control the heat and water flow directly, but it is possible to influence them indirectly. The heat production can be indirectly controlled using ambient temperature $\Ta$, and valve positions $\val$ affect flow:
\begin{align}
    \label{eq_control_inputs}
    \dotQhp(t) = f(\Ta, \cdot),  \hspace{0.3cm} \flow = g(\val, \cdot), \hspace{0.2cm} \val \in \mathbb{R}^N, \hspace{0.2cm} \Ta \in \mathbb{R}
\end{align}
The $(\cdot,\cdot)$ notation indicates that heat and water flow are not only functions of ambient temperature and valve positions, but other factors too.
\subsection{Control hierarchy}
\begin{figure}[h]
	\centering
	\includegraphics[width=0.9\columnwidth]{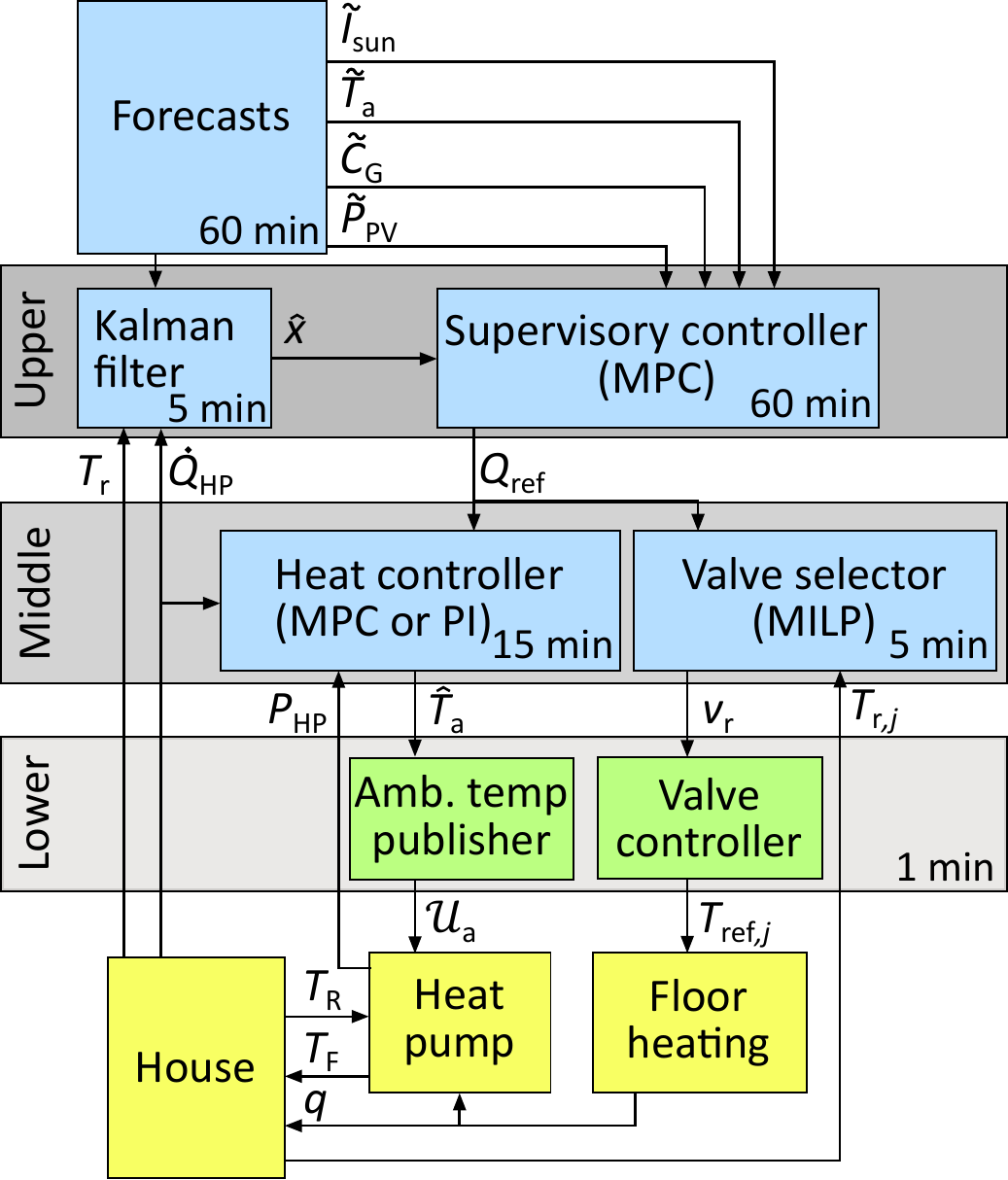}
	\caption{Shows the control diagram with signals. Blue indicates computations conducted remotely and green indicates onsite units and yellow the physical components. The grey boxes contain the control layers in the hierarchical control structure. The update period is shown in the modules.}
	\label{fig_control_diagram}
\end{figure}
The control concept comprises three control levels, see \figref{fig_control_diagram}. The upper layer contains the supervisory controller that is aware of energy assets connected to the system as well as important externalities such as weather and electricity prices. It treats the energy assets as objects with properties which can be utilized for optimal control. A key feature of the supervisory controller is that it knows what the energy assets can do, and why they should do it, but not how to make them do it. The middle layer is tasked with tracking the heat reference, delivered by the supervisory controller, and distributing the heat to appropriate rooms. This layer knows how to deliver the demanded energy, but not why it does it. Based on the heat reference and room temperatures, the valve selector chooses the valves to be opened in order to provide a flow, which works as an operating point for the heat controller, and to transport the heat to the rooms that need it the most. The heat controller follows the heat reference by providing an artificial ambient temperature to the ambient temperature publisher to indirectly control the compressor speed.

The lowest layer handles the interface between the control signal and the actual hardware. The ambient temperature publisher translates the artificial ambient temperature provided by the heat controller to a voltage which emulates the outdoor temperature sensors output at given temperature. The valve controller translates the valve-selection into room temperature references designed to force circuits open or closed.
\subsection{Supervisory controller}
\begin{figure}[H]
	\centering
	\includegraphics[width=0.6\columnwidth]{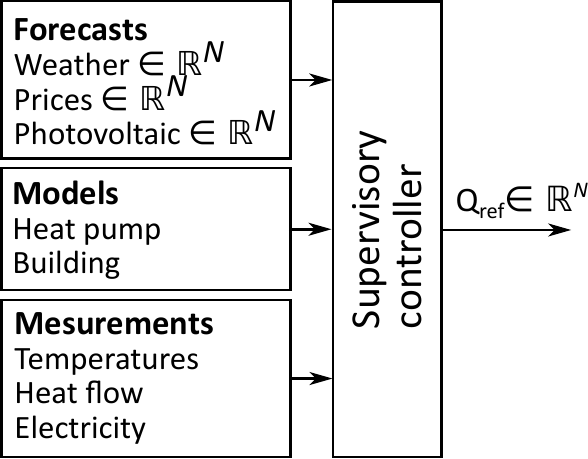}
	\caption{Overview of the top layer supervisory controller.}
 	\label{fig_supervisory_controller}
\end{figure}

\figref{fig_supervisory_controller} presents the concept for the supervisory controller in the upper layer. 
The controller relies on three main components, forecasts, models and measurements. Based on these, the controller computes a heat reference (or ``budget''), $\Qref$, which is dispatched to the lower level controllers. The hierarchical structure makes the supervisory controller more flexible than a monolithic structure, since it can calculate the heat reference without concern for how the heat is delivered---it just needs to know at which efficiency and rate the heat can be delivered. The biggest drawback of using heat for the interface is that it needs to be measured, and adding a heat flow sensor to the hydraulic network is costly.

%% file: sections/method/mpc.tex
\subsubsection{Design of supervisory controller}
\label{sec_mpc}
The conceptualized version of the \textit{\miocp} (\MIOCP) at the core of the \textit{\mimpc} (\MIMPC)\ is seen in equation system \eqref{eq_concept_miocp} on a form which describes the functionality of the cost and various constraints rather than the implementation. Note that \eqref{eq_concept_miocp} contains two sub-versions decided by the indicator variable $\delta_{\op}$. It must be stressed that the value of \deltaOp\ is chosen before implementing the problem, it is not an optimization variable (in this case study: $\delta_{\op}=1$). The difference between the two versions is the \hp\ \efficiencyModel.
\setlength{\jot}{7pt}
\vspace{-0mm}
\begin{subequations}
	\label{eq_concept_miocp}
	\begin{align}
		&J(u_\op^*) = \min_{u_\op} \fgrid(\Pg) + \fcoTwo(\Pg) + \fcomf(\Tr, \Tref) \\
	& u_\op = \begin{cases} (\Php, \delta_{\hp}, \slackVar) & \delta_\op = 1 \hspace{0.1cm} \\ (\dotQhp, \delta_{\hp}, \slackVar) & \delta_\op = 0 \end{cases} \label{subeq_concept_input_var}\\
	 &u_\op \in \realv{\Nhor} \times \{0,1\}^\Nhor \times \realv{\Nslack}_+\\
		&\text{s.t.} \nonumber\\
		&  \Pg = \Php - \forPpv + \Papp \label{subeq_elec_balance}\\
		& \xv_{\iterOp+1} = \Am\xv_{\iterOp} + \Bm \dotQhpi{\iterOp} + \Em \dv_\iterOp, \hspace{0.5cm} \Tri{\iterOp} = \Cm \xv_{\iterOp} \label{subeq_conccept_dynamics}\\
		&\text{if} \hspace{2mm} \delta_{\op} = 1 \hspace{2mm} \text{then} \hspace{2mm} \dotQhpi{\iterOp} = \fhpi{\dotQ}(\Phpi{\iterOp}, \Tai{\iterOp})\deltahpi{i} \label{subeq_concept_if_Q} \\
		&\text{if} \hspace{2mm} \delta_{\op} = 0 \hspace{2mm} \text{then} \hspace{2mm} \Phpi{\iterOp} = \fhpi{\pow}(\dotQhpi{\iterOp}, \Tai{\iterOp})\deltahpi{i} \label{subeq_concept_if_P} \\
		& \Phpi{\iterOp} \in  \begin{cases} [\Phpmin,\Phpmax] & \delta_{\hp,\iterOp} = 1 \\ 0 & \delta_{\hp,\iterOp} = 0 \end{cases}\label{subeq_concept_php_range} \\
		&\Delta\delta_{\hp,\iterOp} = \delta_{\hp,\iterOp} - \delta_{\hp,\iterOp-1} \hspace{0.3cm} \Delta \Phpi{\iterOp} = \Phpi{\iterOp} - \Phpi{\iterOp-1}\\
		&\Delta \Phpi{\iterOp} \in  \begin{cases} [\Delta\Phpmin,\Delta\Phpmax] & \delta_{\hp,  \iterOp} = 1 \\ (-\infty,\Delta\Phpmax] & \delta_{\hp,  \iterOp} = 0 \end{cases} \label{subeq_concept_delta_php_set}\\
 		&\Delta\delta_{\hp,\iterOp} = -1 \implies \delta_{\hp,\iterOp+1},\dots,\delta_{\hp,\iterOp+M-1} = 0 \label{subeq_concept_offtime}
	\end{align}
\end{subequations}
The cost function is the sum of three functions. First, a linear term, $\fgrid(\Pg)$, describing the differentiated cost of either importing from or exporting to the electricity grid. The input is consumed electricity from the grid, \Pg, with positive values indicating import. The prices for buying and selling to the grid are given as $\priceElec^+ > 0$  and $\priceElec^- > 0$, respectively. The second term is a self-imposed $\coTwo$-tax. The third is the comfort term which punishes deviations from the desired temperature. Slack variables are used to ensure feasibility. 
Together the terms make out a convex cost-function.
 
The constraint \eqref{subeq_elec_balance} describes the electricity balance were the amount of electricity bought from the grid (\grid) is calculated. Constraint \eqref{subeq_conccept_dynamics} describes the linear dynamics of the house. Constraints \eqref{subeq_concept_if_Q}-\eqref{subeq_concept_offtime} models the properties of the \hp\ presented in Section \ref{sec_hp_object}. Constraints \eqref{subeq_concept_if_Q} and \eqref{subeq_concept_if_P} both describes the \hp\ efficiency, but only one is active dependent on the initial choice of \deltaOp. Constraint \eqref{subeq_concept_php_range} describes the piece-wise function where the compressor either is off, or operating in the range $[\Phpmin,\Phpmax]$. The constraint \eqref{subeq_concept_delta_php_set} limits the rate off change between control periods. To meet the requirement that the \hp\ can be turned off from any operational state, the down rate is set to $-\infty$ when $\deltahpi{\iterOp} = 0$. Last \eqref{subeq_concept_offtime} forces the \hp\ to stay turned off for minimum $M$ sample times. Having described the functionality of the \OP\, the next part focuses on implementation aspects.

The guiding principle for the implementation is that the structure of the problem is convex if the problem is relaxed, meaning that if integer variables are replaced with continuous ones, the problem is convex. The cost function from equation system\eqref{eq_concept_miocp} is implemented as:
\vspace{-0mm}
\begin{align}
    &J(u_\op) = \priceElec^{-\tra}\Pg + \Delta\priceElec^{+\tra} \Pg^+ + \zcomf + c_\slack^T \slackVar
\end{align}
Here the auxiliary variables $\Pgplus, \zcomf \in \realv{\Nwin}_+$ are introduced. The variable $\Pgplus$ is defined as entry-wise $\max(0, \Pg)$ and \zcomf\ has to be larger than any competing comfort constraints. 
The vector $\Delta\priceElec^{+} = \priceElec^{+} - \priceElec^{-} > 0$ describes the positive difference between buying price and selling price. Note that the buying price needs to be higher than the selling price, otherwise the solution to the \OP\ entails buying excessive amounts of electricity just to sell it again in the same instance. 
The auxiliary variable \zcomf\ encodes the expression $\max(\fcomfi{1}(\Tr, \Tref), \cdots, \fcomfi{\Ncomf}(\Tr, \Tref))$
where $\fcomfi{i}(\Tr, \Tref)$ with $i \in \{1,...,\Ncomf\}$ is either an affine or quadratic positive definite function. This formulation gives room for skewed functions which can for instance penalize either over- or under-heating. Note that the artificial \coTwoTax\ term is not missing, it is merely incorporated into the buying price as described in Section \ref{sec_price_model}.

The \hp\ \efficiencyModel\ in either \eqref{subeq_concept_if_P} or \eqref{subeq_concept_if_Q} is implemented using the known \mld\ technique from \cite{bemporad_control_1999} where an auxiliary variable is introduced $\zhpi{i}$ to either be zero of mirror the value of the function dependent on $\deltahpi{i}$. To preserve convexity of the input set, only an inequality is used instead of the original equality seen in \eqref{subeq_concept_if_P}. if $\deltaOp = 0$ then $\Phpi{\iterOp} \geq \fhpi{\pow}(\dotQhpi{\iterOp}, \Tai{\iterOp})$ and if $\deltaOp = 0$ then $\dotQhpi{\iterOp} \leq \fhpi{\dotQ}(\Phpi{\iterOp}, \Tai{\iterOp})\deltahpi{i}$. The structure of the problem forces the solution onto the curve emulating the equality constraint. When $\deltaOp = 1$, there are a few cases where $\dotQhp$ deviates from the curve to avoid the cost of overheating. To avoid this an equality constraint can be implemented with the added computational cost. 
The constraint in \eqref{subeq_concept_php_range} is implemented as e.g. in \cite{kuboth_economic_2019,lee_mixed-integer_2019,parisio_model_2014}, so is the constraint in \eqref{subeq_concept_delta_php_set}. The down-time model constraint in \eqref{subeq_concept_offtime} can be implemented as shown in \cite{parisio_model_2014}.
The problem can be summed up to
\vspace{-0mm}
\begin{subequations}
    \label{eq_miocp}
    \begin{align}
        &\min_{\uv \in \realv{m \times N}, } J(\xv_0, \uv)\\
        &\text{s.t.}\\
        &\SSdx{k}, \hspace{0.2cm} \xv_{k} \in  \mathcal{X},\hspace{.2cm}\xv_0 = \xv(t)\\
        &\yv_{1,k} = \Cm\xv_k + \Dm\uv_k, \hspace{0.2cm} \uv_{k} \in  \mathcal{U} \\
        &\yv_{2,k} \geq \textbf{f}_\text{convex}(\xv_k, \uv_k) \hspace{0.6cm} \yv_{3,k} \leq \textbf{f}_\text{concave}(\xv_k, \uv_k)
    \end{align}
\end{subequations}
where $J$ is the convex cost function. Section \ref{sec_model} details the models that specify the \MPC\ formulation given here.

%% file: sections/method/valve_control.tex
\subsection{Valve selector}
\vspace{-2mm}
The valve selector, or dispatcher, is a mixed integer linear programming problem tasked with providing the flow, $\flow$, as requested, $\flow_\reference$, and distribute the water to the most suitable rooms. Note that the Valve selector is only reactive control. The optimization problem is
\label{sec_sub_control}
\newcommand{\valVSpace}{\vspace{0mm}}
\begin{subequations}
	\label{eq_flow_op}
	\begin{align}
		&\min_{\val \in \{0,1\}^M}  J(\val)=\min_{\val \in \{0,1\}^M} c_{\flow} \Vert \flow_\reference - \flow \Vert^2_2 + \priceComf^\tra \val \label{eq_cost}\valVSpace\\
		&\text{s.t.}\valVSpace\\
		& \nomflow = \sum_{i = 1}^{M} \val_i\nomflow_i, \label{constraint_q}\valVSpace\\
		&\flow = c_0 + c_1\nomflow + c_2\nomflow^2 \label{constraint_q_poly}\valVSpace\\
		& \flow \geq \flow_{\min}  \label{constraint_min_q},\valVSpace\\
		&  r_m \geq \sum_{i = 1}^{M} \max(\val_{0,i}-\val_{i}, 0), \label{constraint_limit_closed}\valVSpace\\
		&\val_j = 1, \hspace{1cm} \forall j \in \mathcal{J}.	\label{constraint_force_open}
	\end{align}
\end{subequations}
The cost function is a trade-off between following the flow reference and delivering the heat to the right rooms. The two terms are weighted by $c_{\flow} \in \mathbb{R}_+$ and $\priceComf \in \realv{\Nroom}$. The comfort cost is pre-calculated as $\priceComfi{i} = a(\Tri{i} - \Trefi{i})$ with $a > 0$ such that cold rooms get priority. The expressions \eqref{constraint_q} and \eqref{constraint_q_poly} describe the flow as a function of valve configuration. In \eqref{constraint_q} the flow is a sum of contributions, but since the flow saturates as more valves open a second order polynomial is used in \eqref{constraint_q_poly} to model this effect. The term $\nomflow^2$ seems to deliver a range of square and \bilinear\ terms, which is inconsistent with \MILP. Luckily, $\val_i$ is binary, meaning that $\val_i \val_j$ is an AND statement ($v_i \land v_j$) can be encoded by \mld (\MLD) as a linear inequality \cite{bemporad_control_1999}. The squared terms are unproblematic since $\val_i^2 = \val_i$. Encoding the binary polynomial has a cost in form of added binary auxiliary variables. The added number of binary variables is ${M \choose 2}$, which in this case is 55, making a total of 66 variables. Constraint \eqref{constraint_min_q} forces a minimum flow, \eqref{constraint_limit_closed} limits the number of valves that can be closed in one iteration and \eqref{constraint_force_open} forces circuits open which belong to too cold rooms.

%% file: sections/method/heat_pump_control.tex
\subsection{Heat controller}
\begin{figure}[h]
	\centering
	\includegraphics[width=0.8\columnwidth]{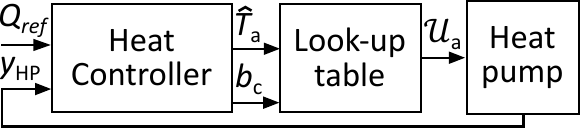}
	\caption{Shows the heat controller with feedback}
	\label{fig_heat_control}
\end{figure}

The heat controller, placed in the middle layer, is required to deliver heat, \Qhp, according to the heat reference. Further, it suppresses the compressor in periods with no demand and is responsible for timing the \hp\ start. The diagram is seen in \figref{fig_heat_control}.
The signals are: reference vector $\Qref \in \realv{N}$, artificial ambient temperature, \artTa, the voltage representing the said ambient temperature, $\Volta$, the measurement vector, $y_\hp = \begin{bmatrix} \dotQhp & \Php \end{bmatrix}$, and the binary compressor blocking signal, $b_\text{c}$. During the test period a PID-controller and a short horizon \MPC\ were tested. The PID-controller uses the measured heat flow and the reference in regular feedback. The \MPC\ accumulates the delivered heat flow over the hour to match the heat reference given for that hour. Beyond heat control, the two controllers need to handle defrosting periods, \DHW\ production and start delays as mentioned in Section \ref{sec_hp_object}. Defrost periods and \DHW\ production are handled by detecting the event and setting the controller to standby-mode. After releasing the compressor block, it takes about 1.5 hour before the heat pump starts, therefore the reference vector is used to remove the blockage a defined time-span before the actual control takes place. More detail is given in a parallel paper in progress.

%% file: sections/method/model.tex
\section{Models and parameter identification}
\label{sec_model}
This section presents the model and the subsequent parameter identification for each module that is included in the \MIOCP.
 Subsections \ref{sec_house_model}-\ref{sec_pv_model} presents models used directly for control while subsection \ref{sec_price_model} contains the price model which is used both for guiding the price-aware controller and evaluation. Finally, subsection \ref{sec_validation} describes how benchmark data from previous heating season, where the baseline controller was running, is used to evaluate the proposed controller.

%% file: sections/method/house_model.tex
\subsection{Single-zone lumped parameter house model}
\label{sec_house_model}
 The single-zone house model, seen in \eqref{eq_heating_model_2order}, has, as argued in \cite{killian_comprehensive_2018}, two dynamic states which describe an averaged room (\Tr) and floor temperature (\Tf). The reason for modelling using only a single zone is given in \cite{vogler-finck_inverse_2019}. Here the affects caused by position of doors and air stratification led the authors to conclude that a single-zone model is as useful as multi-zone model for \MPC. In \cite{amato_dual-zone_2023} a volume weighted average temperature is used since only the central heat meter is available. The purpose of the model is to make the \MPC\ responsive to the impact of high sun intensity, ambient temperature and heat created by household appliances. These three aspects should be included if a forecast is available, otherwise they can be omitted at the cost of increased uncertainty. In this work the forecast for heat produced by household appliances and occupation is left out. The reason is partly technical, but also driven by privacy concerns.
\begin{subequations}
	\label{eq_heating_model_2order}
	\begin{align}
		\Cr\dotTr &= \Ur \left(\Tf - \Tr\right) + \Uamb \left( \Ta - \Tr \right) + \dotQsun  \label{subeq_room}\\
		\Cf\dotTf &=  \Ur \left( \Tr - \Tf \right) + \dotQhp \label{subeq_floor}
	\end{align}
\end{subequations}
The control input to the model is heat flow $\dotQhp$ measured over the floor heating system. The two-state formulation allows for estimating the overall heat capacity of the building through $\Cf$ and to capture the rapid air temperature changes, caused by sun radiation, in $\Cr$. The state space formulation is given as,
\newcommand{\hModelSpace}{\hspace{0.2cm}}
\begin{subequations}
\label{eq_all_ss}
\begin{align}
	\label{eq_ss_model_final}
	&\dot{\xv}(t) = \Am\xv(t) + \Bm \uv(t) + \Em \dv(t)\\
	&\yv(t) = \Cm \xv(t) \hspace{1cm} \Cm = \begin{bmatrix} 1 & 0\end{bmatrix} 
\end{align}
\begin{align}
	\label{eq_ss_matrices}
	&\Am = \Ccal^{-1}\mathcal{U} \hspace{1cm} \Bm = \Ccal^{-1}\mathcal{B} \hspace{1cm} \Em = \Ccal^{-1}\mathcal{E}\\
	&\Ccal = \begin{bmatrix} \Cr & 0 \\ 0 & \Cf\ \end{bmatrix} \hModelSpace \Ucal = \begin{bmatrix} -\left(\Ur + \Uamb\right) & \Ur \\ \Ur & -\Ur  \end{bmatrix} \hModelSpace 
	\Bcal = \begin{bmatrix} 0 \\ 1  \end{bmatrix}\\ &\Ecal = \begin{bmatrix}  \Uamb & \sunParam{1} & \sunParam{2}\\ 0 & 0 & 0 \end{bmatrix} \hModelSpace  \xv = \begin{bmatrix} \Tr \\ \Tf  \end{bmatrix} \hModelSpace \uv = \begin{bmatrix}  \dotQhp  \end{bmatrix} \hModelSpace \dv = \begin{bmatrix}  \Ta \\ \forIsun  \\ \forIsundir\end{bmatrix} \label{eq_state_space_end}
\end{align}
\end{subequations}
The input is total heat flow, \dotQhp, and the disturbances are ambient temperature, direct sun irradiation. The common indoor temperature is an area weighted average of all room temperatures,
\begin{align}
    \label{eq_common_Tr}
    \Tr = \frac{\Arj{1}\Tri{1} + \cdots + \Arj{\numRooms}\Tri{\numRooms}}{\Arj{1} + \cdots + \Arj{\numRooms}} 
\end{align}
The power from sun radiation can be estimated in many ways, but is here chosen to be:
\begin{align}
    \dotQsun &= \sunParam{1} \forIsun + \sunParam{2}\forIsundir\\
    \forIsun &= \forIsundir (1 -\forCloudCover)
\end{align}
where \forIsundir\ [\si{\watt\per\meter\squared}] is direct sun and $\forCloudCover$ is the fraction of cloud cover. This particular formulation gives short but intense bursts of sunlight which can capture the rapid increases in room temperature seen in the data. A similar sun radian term can be added to equation \eqref{subeq_floor}, but note that the coefficients of matrix $\Em$ need to be positive to keep the model grounded in physics.

The model is discretized using \zoh\ (\ZOH) discretization. Figures of the parameter fits are shown in Appendix \ref{app_house_model}.

%% file: sections/method/state_est.tex
\subsection{State estimation of \Tr\ and \Tf}
Since the virtual average floor temperature, used in the \MPC, is not measured a \lkf\ (\LKF) is used to estimate the state at sample time $k$. The \LKF\ is updated each 5 min.
The model in \eqref{eq_all_ss} is observable for any $\Ur, \Uamb > 0$.
\begin{align}
    \mathcal{O} = \begin{bmatrix} \Cm \\ \Cm\Am \end{bmatrix} \ = \begin{bmatrix} 1 & 0 \\ -\left(\Ur + \Uamb\right) & \Ur \end{bmatrix}
\end{align}
This is the case even if the matrices \Am, \Bm\ and \Em\ have been found using a black box method, since the parameter in eq. system \eqref{eq_all_ss} can be solved for, if \dotQhp\ is known.

%% file: sections/method/heat_pump_model.tex
\subsection{Air-to-water heat pump efficiency model}

In order to inform the supervisory controller on the efficiency of the \hp\ a relation between heat production and electricity consumption is formulated. It is inspired by the formulations provided in \cite{wimmer_regelung_2004,baumann_experimental_2023,burger_whole-year_2020,odukomaiya_addressing_2021,yang_model_2022,yang_model_2022-2, dettorre_model_2019}\footnote{Although, all formulations cited here are based on the non-ideality factor multiplied with the Carnot \cop\, they all differ with respect to choice of variables.}, where the \hp\ efficiency is provided by the Carnot coefficient of performance, $\copcarnot = \frac{\T_\hot}{\T_\hot-\T_\cold}$, and a non-ideality factor (also called efficiency factor), \etahp:
\begin{align}
    \label{eq_cop_basic}
    \dotQhp &= \etahp(\Php) \copcarnot \Php \nonumber \\
    &= \cophp \Php
\end{align}
with \cophp\ being the overall efficiency for a given \hp. The expression for heat as a function of electricity (direct way) is denoted as $\fhpi{\dotQ}$ and the reverse way where electricity is calculated from heat is  $\fhpi{\pow}$.
\subsubsection{Heat as a function of electricity: $\fhpi{\dotQ}$}
The expression chosen for the \cophp\ is:
\begin{align}
	\label{eq_regr_cop_2}
	&\cophp = \frac{k}{\Php } +
	\left(\frac{k_0}{\Php } + k_1 + k_2\Php\right)\copcarnot
\end{align}
with
\begin{align}
\copcarnot \equiv \frac{\TF+273.15}{\TF-\Ta}
\end{align}
and the requirement that the coefficient $k_2$ is negative and the forward temperature is constant \nomTF. The reason for this choice is that the heat function $\fhp(\Php)$ contains a second order polynomial when the power \Php\ is multiplied onto \eqref{eq_regr_cop_2}:
\begin{align}
	\label{eq_regr_Q_2}
	\dotQhp = k + (k_0 + k_1\Php + k_2\Php^2)\copcarnot
\end{align}
and taking the second order partial derivative of \eqref{eq_regr_Q_2}
with respect to \Php\ shows that 
\begin{align}
	\label{eq_hessian_Q_2}
	\frac{\partial^2\dotQhp}{\partial^2 \Php} = k_2\copcarnot < 0
\end{align}
implying that if all other variables are constants then \eqref{eq_regr_Q_2} is concave.
\subsubsection{Electricity as a function of heat: $\fhpi{\pow}$}
The inverse formulation, seen in \eqref{eq_regr_P_2}, where electricity is the dependent variable, is convex if the coefficient $k_2 > 0$. 
\begin{align}
	\label{eq_regr_P_2}
	\Php = k + (k_0 + k_1\dotQhp + k_2\dotQhp^2)\frac{1}{\copcarnot}
\end{align}

%% file: sections/method/pv_model.tex
\subsection{Photovoltaic power forecast}
\label{sec_pv_model}
The forecast model for the power output of the photovoltaic panels (\pv) is based on the data from the weather forecast service Yr.no \cite{yr} and measured historical time series of the power output from the \pv. In this work it is chosen to be a regression expression, although the model for the predicted \pv\ output could in principle be any suitable non-linear model (neural network, decision tree, etc.) since the produced electricity is not dependent on any influenceable variables.

%% file: sections/method/price.tex
\subsection{Price model}
The hourly price models for buying and selling electricity from/to the grid is given in \eqref{eq_price_buy} and \eqref{eq_price_sell}, respectively. The models are used in the supervisory controller and for evaluation. 
The price for buying electricity is
\begin{align}
    \price^{+}_{\text{excl. \VAT}} &= \priceSpot + \priceTariff + w_{\coTwo} \priceCoTwo + \priceTransport \nonumber\\ 
    \priceElecBuy &=  \price^{+}_\text{excl. VAT} + 0.25\price^{+}_{\text{excl. \VAT}}
    \label{eq_price_buy}
\end{align}
where the spot price, \priceSpot, distribution tariff, \priceTariff, transport tariff, \priceTransport, and are given in [\si{\text{\euro}\per\kWh}]. The self-imposed artificial \coTwoTax, \priceCoTwo, is given in [\si{\text{\euro}\per\kilogram}]. Hence, the variable $w_{\coTwo}$ is the hourly estimated \coTwo\ emission in kg per kWh electricity. The Danish VAT rate is 25\% of the full price and the \tsos\ (\TSOs) tariff is a fixed rate of \euro 0.02. The selling price model is
\begin{align}
    \label{eq_price_sell}
    \priceElecSell = \priceSpot
\end{align}
The distribution tariffs, \priceTariff, chosen for the test are based on future signaled prices for January 1st, 2023 in Denmark. The exact tariffs vary between Distribution Systems Operators (\DSOs s), but the pattern is low prices at night, a higher daily price with a sharp increase in the cooking peak. The chosen model is inspired by \cite{net_tariffs_cerius}.
\label{sec_price_model}
\begin{align}
    \priceTariff =  \begin{cases} 
        0.027 \text{\euro} & t \in {[00.00,06.00)} \\ 
        0.081 \text{\euro} & t \in {[06.00,17.00),[21.00,00.00)} \\
        0.26 \text{\euro} & t \in {[17.00,21.00)}
    \end{cases}
\end{align}
It is worth noting that the tariffs need to be realistic, since the choice of values has a large impact on savings potential. If an unrealistic price of \euro 10 is used for the \cookingPeak\ instead of \euro 0.26, the price-aware controller shuts the \hp\ off in this period and gains and unfair advantage over the price-unaware.

The second part of the price model regards \pv\ produced electricity and the impact the \hp\ has on self-consumption. In the test house the electricity is phase-metered, but the exact per phase import and export is unknown since the numbers are aggregated and stored on hourly basis. Since the data is aggregated, the meter is instead treated as a summation meter with one-hour reporting. The netting interval is unknown even though it is important for the measure of import and export, as shown in \cite{ziras_effect_2021}. The available signals are hourly import, $\Eimp(k)$, hourly export, $\Eexp(k)$, hourly production from the \pv, $\Epv$ and consumption from \hp, \Ehp(k). The difference between export and import, seen in \eqref{eq_net_import}, is the net import, $\Delta\Eg(k)$, which is the billable amount. For notational purposes the hour indicator $k$ is implied hence on.
\begin{align}
    \label{eq_net_import}
    &\Delta\Eg = \Eimp - \Eexp
\end{align}
The sun power corrected cost associated with running the \hp\ is then:
\vspace{-1mm}
\begin{align}
    \Ehp^{*} =  \begin{cases} 
        0 & \Delta\Eg \leq 0 \\ 
        \min\left( \Ehp, \Delta\Eg\right) & \Delta\Eg > 0
    \end{cases}
\end{align}
The same amount of available \pv\ produced solar power and consumption is imposed on similar/comparable days (definition in section \ref{sec_validation}), which are used in the controller evaluation. The consumption of similar days is corrected using the difference in \hp\ consumption: 
\vspace{-1mm}
\begin{align}
    \label{eq_hp_diff}
    \Delta\Ehp =  \Ehpcmp - \Ehpexp
\end{align}
with \Ehpexp\ being the \hp\ consumption for the experiment and \Ehpcmp the similar day. The virtual net import increases when the \hp\ consumes more in hour $k$ and vice versa, as seen in \eqref{eq_Eq_corr}.
\vspace{-1mm}
\begin{align}
    \label{eq_Eq_corr}
    \Delta\Eg^{*}(k)  =  \Delta\Eg + \Delta\Ehp(k)
\end{align}
The corrected net import for similar/comparison days is:
\begin{align}
    \Ehp^{*} =  \begin{cases} 
        0 & \Delta\Eg^{*} \leq 0 \\ 
        \min\left( \Ehpcmp, \Delta\Eg^{*}\right) & \Delta\Eg^{*} > 0
    \end{cases}
\end{align}
The idea behind this mode of calculating the \hp\ consumption is that other appliances use the self-produced electricity too, and the \hp\ should ideally consume less than the excess capacity.


%% file: sections/method/validation.tex
\subsection{Evaluation procedure}
\label{sec_validation}
\newcommand{\iterDayCmp}{j}
\newcommand{\iterDayExp}{i}
\newcommand{\iterDay}{i}
The objective of the evaluation procedure is to answer whether the new price- and forecast-aware controller saves money when compared to the existing benchmark controller described in Section \ref{sec_test_house}. The key performance indicator is daily cost given the weather conditions. It is inherently difficult to benchmark and validate the performance of a controller operating in a complex environment with many uncontrollable external factors such as weather and occupant activities. Further, the long time-constants play a significant role by demanding long test periods. Ideally, the benchmark  and \MPC-controller should be run in parallel on exact copies of the same building placed at the same location, with occupants doing the same activities. Although, some buildings support such circumstances, this can obviously not be asked of the occupants. Instead, a benchmark \dataSet\ from the same house is used for the evaluation. The benchmark \dataSet\ is based on data collected from the former heating period (2021-2022) where the original benchmark controller was operating. The data is sorted into full days creating a collection of comparison days $\daySetCmp$, seen in \eqref{eq_days}, from which appropriate subsets can be selected. The daily generated data on set form is:
\newcommand{\iterDayOne}{n}
\begin{align}
    \label{eq_days}
    &\mathcal{D}_\compare = \left\{day^\iterDayOne = \left(\Eg^\iterDayOne, \Ta^\iterDayOne, \priceElecBuySeti{\iterDayOne},\Epv^\iterDayOne  \right)\right\}\\
    &\iterDayOne = 1,\dots,\Ncmp, \hspace{4mm} \Eg^\iterDayOne, \Ta^\iterDayOne, \priceElecBuySeti{\iterDayOne},\Epv^\iterDayOne \in \realv{\Nday}
\end{align}
with $\Eg^\iterDay$ and $\Epv^\iterDay$ being the electricity consumption from grid and production from \pv\ in \si{\kWh} during day $i$, respectively, $\Ta^\iterDay$ the ambient temperature, $\priceElecBuySeti{\iterDay}$ the hourly electricity price for day $i$, and $\Nday = 24$. Note that benchmark days where the system has been manipulated or a significant amount of data is missing are dropped to minimise pollution of the results. A similar data collection, $\daySetExp$, is generated from the experiment period. 
The \MPC-controller is evaluated daily by comparing the operation cost of day $\iterDayExp$ to a subset of benchmark days, $\daySetCmp^\iterDayExp \subset \daySetCmp$, drawn from the full benchmark \dataSet. The subset, $\daySetCmp^\iterDayExp$, is drawn according to the following rule:
\begin{align}
    \label{eq_subset_cmp_days}
    \mathcal{D}_{\compare}^\iterDayExp &= \{day \hspace{1mm}\vert\hspace{1mm}  \nonumber\\ &-\Delta\avgTai{\dn} \leq \avgTaCmp - \avgTaExp^\iterDayExp \leq \Delta\avgTai{\up}, \nonumber\\
    &-\Delta\Epvi{\dn}  \leq \Sigma \EpvCmp-\Sigma\EpvExp^\iterDayExp \leq \Delta\Epvi{\up}, \nonumber \\
    & \avgTaCmp,\Sigma \EpvCmp \in day \in \daySetCmp \}
\end{align}
with $\avgTa$, $\Sigma \Epv$ being average ambient temperature and accumulated electricity production from \pv, respectively. The constants $\Delta\avgTai{\dn}$ and $\Delta\avgTai{\up}$ are the down- and up-search range for ambient temperature, respectively. Similar, $\Delta\Epvi{\dn}$, $\Delta\Epvi{\up}$ makes out the search-range for accumulated electricity produced by the \pv. Here the \pv\ is used as an indicator for sun radiation. This is not a perfect indicator, since the sun altitude and intensity vary with the seasons, thereby creating a bias. However, it is found to be a good indicator for dealing with cloud conditions on-site, since it directly measures the level of shadow on the building. With ambient temperature and sun irradiation accounted for, factors such as occupant behavior and previous day heating patterns are left out. This undeniably causes noise, making the electricity consumption of the \hp\ distribute randomly for any given day. To decrease the influence of the noise, the controller is run over a long period to obtain more consistent results.

We calculate a virtual cost for benchmark day $\iterDayCmp$, with respect to experiment day $\iterDayExp$,
\begin{align}
    \label{eq_cost_comp}
    &\costElecCmp^\iterDayCmp = \sum_{k = 0}^{\Nday} \priceElecBuySeti{\iterDay}(k)\Eg^\iterDayCmp(k)\nonumber \\  &\priceElecBuySeti{\iterDay} \in \dailyExp^\iterDayExp, \hspace{0.4cm} \Eg^\iterDayCmp \in \dailyCmp^\iterDayCmp \in \daySetCmp^\iterDayExp
\end{align}
It simply means that electricity consumption from similar benchmark days are imposed onto the price of the experiment day to calculate the virtual cost. This provides a plausible alternate outcome for the case where the benchmark controller had been running instead. This is done since the benchmark controller is price ignorant and thereby acts independently of the price. This manoeuvre would not be possible if the comparison was between two price-aware controllers. In that case price curves would have to be accounted for as well. The cost of the experiment day $i$, $\costElecExp^\iterDayExp$, is of course calculated using the actual electricity consumption for the day.

%% file: sections/results/results.tex
\input{sections/results/experiment}

\section{Results}
This section presents the results gained from the experiment period. Note that with respect to the cost analysis comfort level 3 ({\color{ref3}\FilledBigTriangleUp}) and 4 ({\color{ref4}\FilledBigTriangleUp}) should be given the highest attention since they both, on average, feature higher indoor temperatures than the benchmark period and therefor provides the best foundation for investigating the savings potential. Comfort level 1 ({\color{ref1}\FilledBigTriangleUp}) and 2 ({\color{ref2}\FilledBigTriangleUp}) are included to provide a broader overview of the challenges related to carrying out the experiment.
\label{sec_results}

\input{sections/results/key_results}
\input{sections/results/temp_comfort}
\input{sections/results/cost}
\input{sections/results/heat_pump_efficiency}
\input{sections/results/production}

%% file: sections/results/experiment.tex
\section{Experiment description}
\label{sec_experiment}
\begin{figure}[ht]
	\centering
	\includegraphics[width=1\columnwidth]{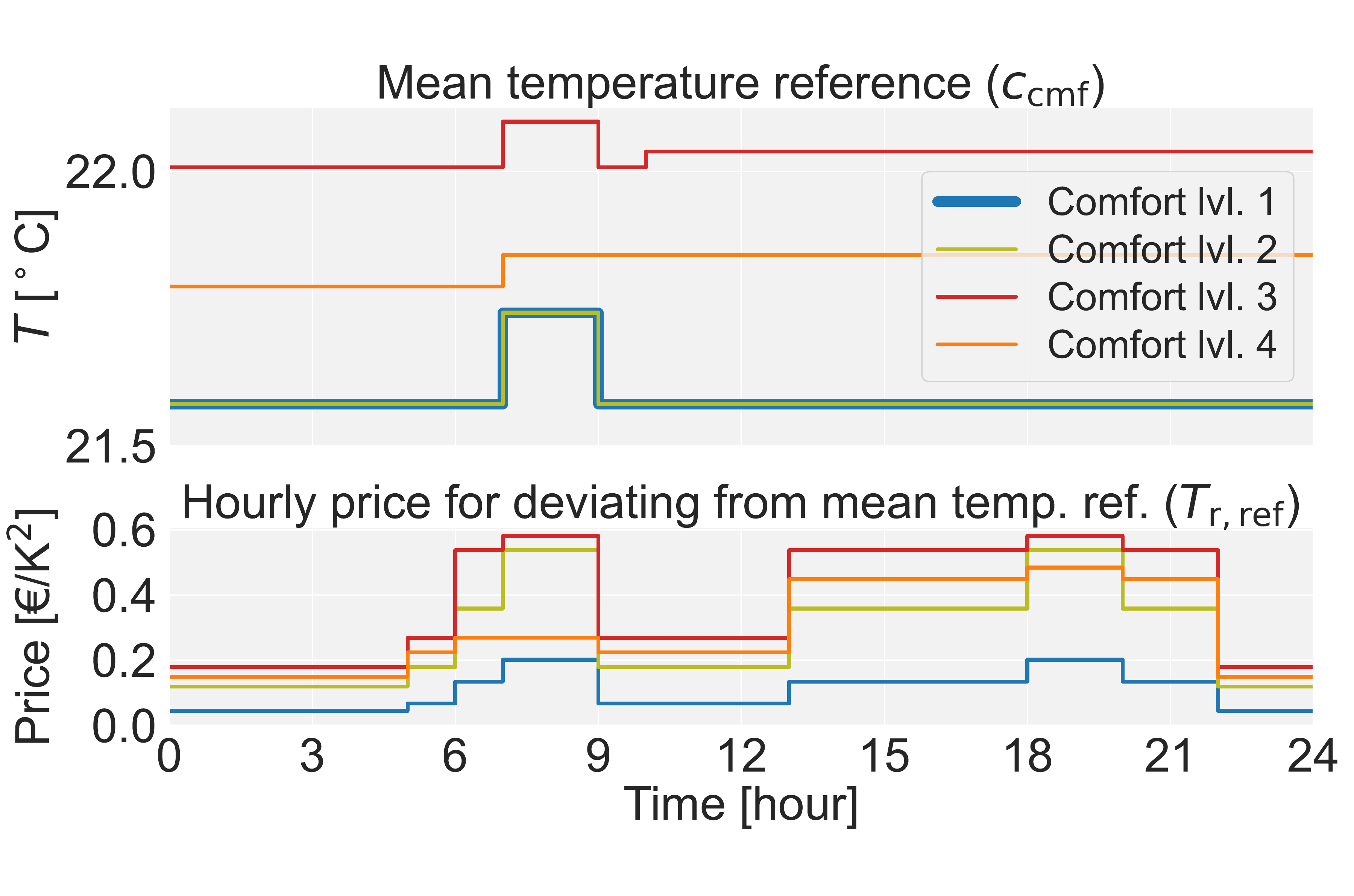}
	\caption{\figUpper Hourly mean temperature reference calculated from the collection of zone references. \figLower Hourly virtual discomfort price. Both vectors are used in equation system \eqref{eq_concept_miocp}.} 
	\label{fig_comfort_cost}
\end{figure}

\begin{figure}[h]
	\centering
	\includegraphics[width=\columnwidth]{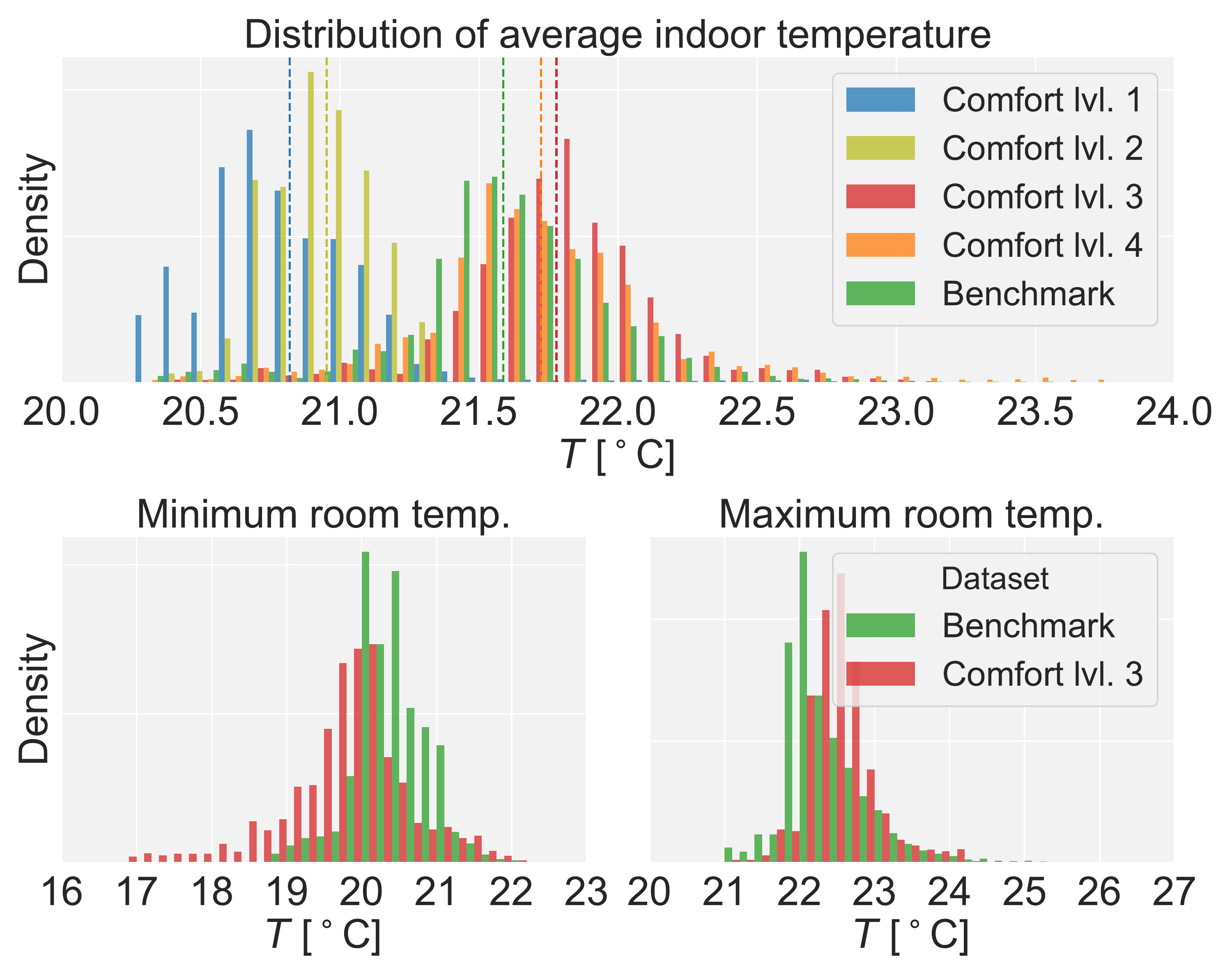}
	\caption{\figUpper\ Distribution of mean indoor temperature compared to the benchmark period. \figLowerLeft\ Histogram of the lowest measured temperature in the rooms. Dashed line shows the mean temperature. \figLowerRight\ Highest measured temperature in the rooms.}
	\label{fig_avg_temp_1}
\end{figure}

\newcommand{\numComfLvlWord}{four}
\newcommand{\numComfLvl}{4}
\newcommand{\TaBound}{0.5}
\newcommand{\TaBoundPerc}{5}
\newcommand{\EpvBound}{2.0}
\newcommand{\EpvBoundPerc}{10}
The experiment was conducted over \numTestDays\ days in the period 2022-11-07 to \expEndDate. During the experiment \numComfLvlWord\ combinations of hourly discomfort cost, $\priceComf \in \realv{24}$, and average room temperature reference levels, $\Tref \in \realv{24}$, were applied, see \figref{fig_comfort_cost}. A pair consisting of a temperature reference and a discomfort cost makes out a comfort level. Each comfort level has a colour assigned  ({\color{ref1}\FilledBigTriangleUp}, {\color{ref2}\FilledBigTriangleUp}, {\color{ref3}\FilledBigTriangleUp}, {\color{ref4}\FilledBigTriangleUp}) corresponding to the colours in \figref{fig_comfort_cost}. The cost and reference are used in the quadratic cost term $\sum_{i=0}^{23} \priceComfi{i} \left(\Tri{i} - \Trefi{i} \right)^2$, where $i$ is the hour. 
Having \numComfLvlWord\ comfort levels is a result of gradually adjusting the overall average indoor temperature to be similar to the one from the benchmark data in order to reduce a variable with respect to the cost analysis. Consistency of indoor temperature was achieved at comfort level 4 as seen in the upper graph in \figref{fig_avg_temp_1} where the average temperature distributions are plotted. The gradual adjustment of the reference and comfort cost was necessary because the resulting room temperature is difficult to predict since it is dependent many factors such as electricity price, heat pump efficiency, heat loss of the house, etc.

\begin{figure}[h]
	\centering
	\includegraphics[width=\columnwidth]{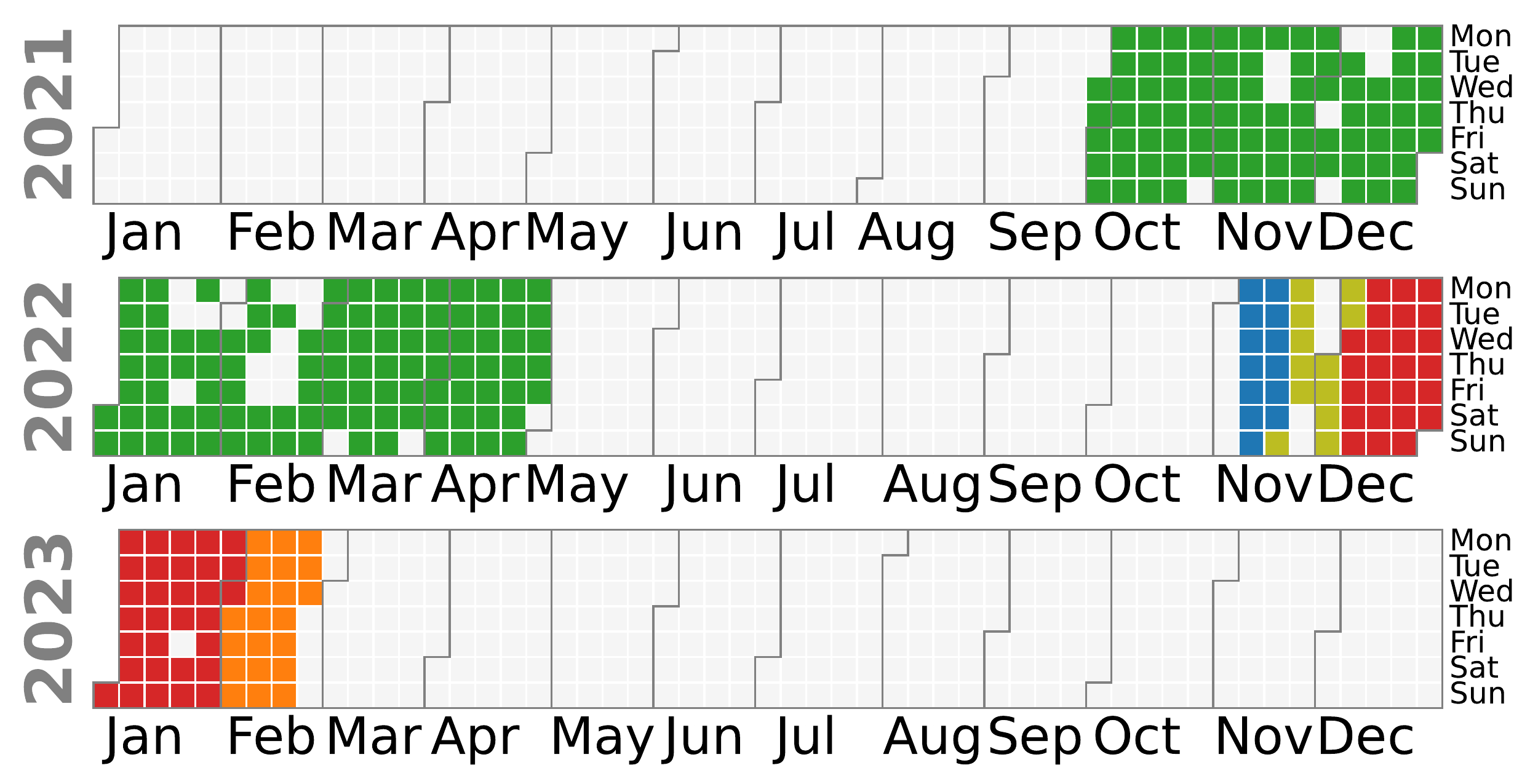}
	\caption{Calendar overview of benchmark data and experiment periods. Green color is benchmark days and other colors are the comfort levels during the experiment.}
	\label{fig_calendar}
\end{figure}

The benchmark and experiment periods are shown in \figref{fig_calendar}.
The benchmark dataset used for comparison consists of 193 days with daily mean ambient temperatures in the range $-2.5$ to $\SI{15}{\degreeCelsius}$ and daily \pv\ production in the range 0 to \SI{39}{\kWh}. The search bound for finding similar benchmark days are 0.5 \degC\ for ambient temperature making it \TaBoundPerc\% of the full range and \SI{\EpvBound}{kWh}\ for daily \pv\ electricity production which is \EpvBoundPerc\% of the full range.

The \hp\ efficiency model used for the entirety of the case study was \eqref{eq_regr_Q_2}. The coefficients were re-calibrated twice during the experiment. First time was after a few weeks of running \MPC\ and second the $27^{\text{th}}$ of January. The building model was changed and refitted on the $27^{\text{th}}$ of January.

The optimization problems are implemented using Casadi \cite{Andersson2019} and solved with the mixed integer non-linear programming solver Bonmin \cite{bonmin}.

%% file: sections/results/key_results.tex

%% file: sections/results/temp_comfort.tex
\subsection{Temperature comfort}
\label{sec_temp_comfort}
The temperature distributions in \figref{fig_avg_temp_1} show that the indoor temperature has not been impacted by the \MPC\ controller providing price-led load shifting. This is particularly clear in the lower left and right plot which shows the distribution of minimum and maximum room temperatures, respectively. The min./max. room temperature are defined as $\min/\max\left(\Tri{1}(t), \dots, \Tri{\numRooms}(t)\right)$. The lower minimum temperature is caused by one room where the reference was set to 19 \degC.

Since the house was occupied throughout the test, the residents were sent a questionnaire about the experienced indoor climate on the $11^\text{th}$ Jan. 2023. The questions and answers can be read in Appendix \ref{app_resident_statement}.

Although each room has an assigned temperature reference, not much attention has been given to individual rooms besides responding to complaints, which  was only necessary once, at comfort level 1. Two rooms, hobby and bedroom, had reference settings at 19 and 21 \degC, respectively, and the rest had 22.5 \degC. The hobby room is partly detached from the rest of the house, and it was thus easy to keep the temperature low. The bedroom could not be kept at 21 \degC, even though the floor heating circuit was seldom on. This shows, as pointed out by \cite{vogler-finck_inverse_2019}, that it is difficult to maintain large discrepancies between room temperatures within a \NZEB.

%% file: sections/results/cost.tex
\subsection{Heating costs and energy consumption}
\label{sec_res_cost}
\begin{figure}[H]
	\centering
	\includegraphics[width=1\columnwidth]{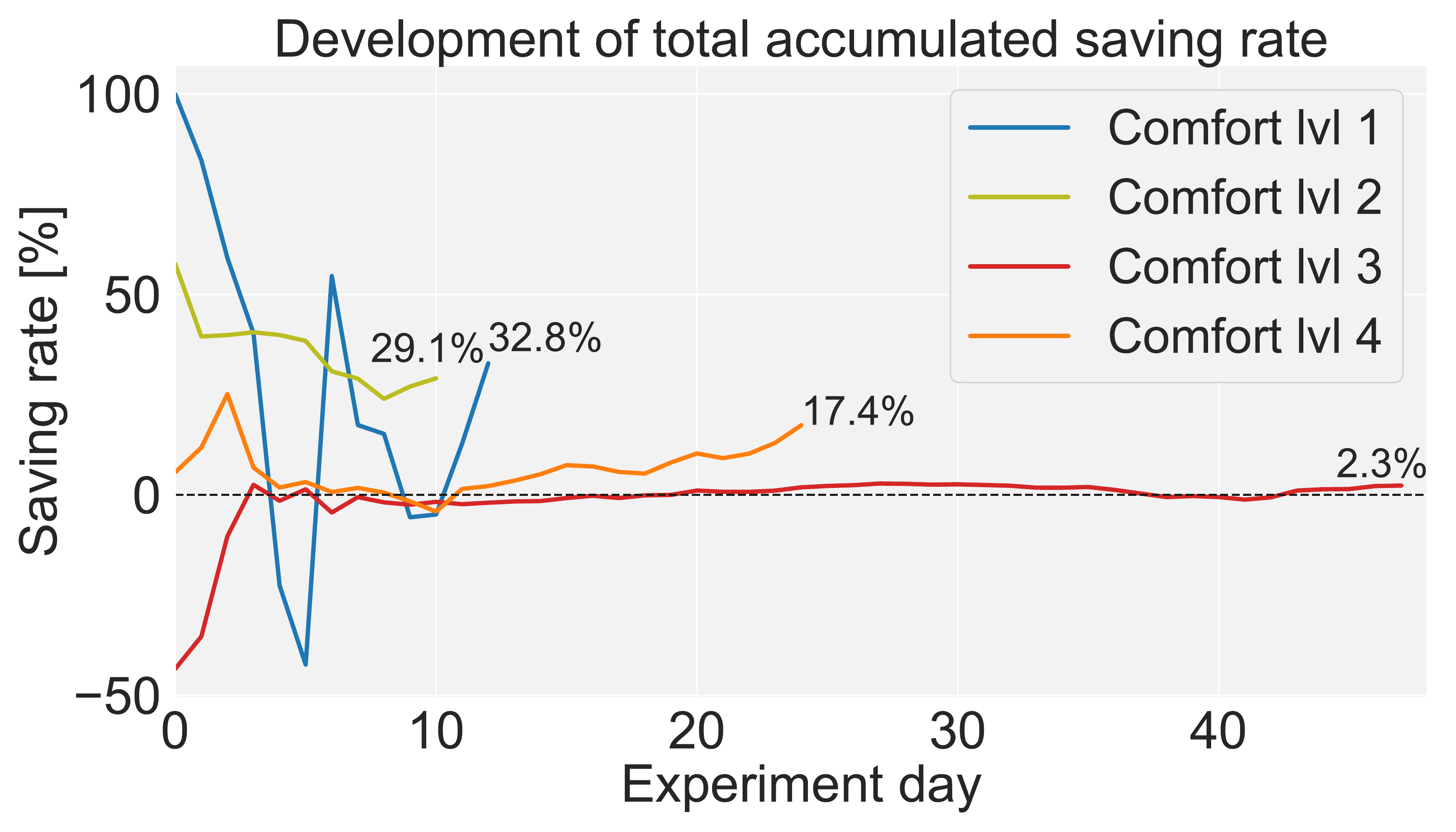}
	\caption{Day-to-day development of the total accumulated \savingRateText\ for each comfort level.}
	\label{fig_savings_development}
\end{figure}

\begin{sidewaysfigure*}
	\centering
	\includegraphics[width=\textwidth]{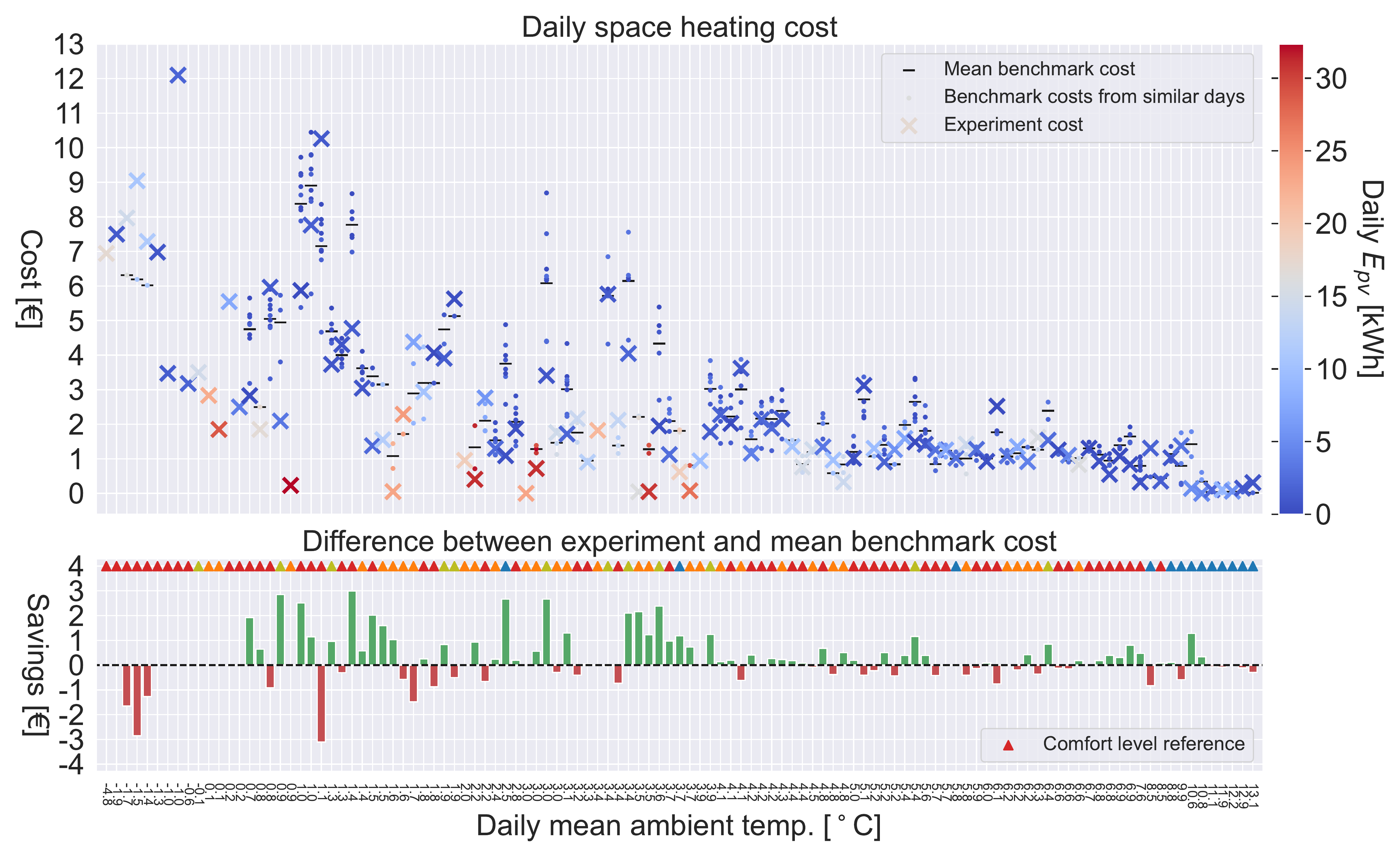}
	\caption{\figUpper Daily space heating cost during the experiment period (crosses) compared with the virtual costs, calculated based on Section \ref{sec_validation}, from similar days (dots). Each experiment day is assigned one column and are sorted according to daily mean temperature. The black horizontal line (-) marks the mean benchmark cost. \figLower Savings between mean benchmark cost and experiment cost. The colors of the triangles refer to the comfort level at given experiment day.}
    \label{fig_cost_compare}
\end{sidewaysfigure*}

This section is dedicated to the investigation of the savings potential.
The section consists of Table \ref{table_cmp_cost}, which sums up the savings accumulated during the test periods and \figref{fig_cost_compare} presents costs with respect to individual days. Note that a row in each table has been dedicated the combined analysis of comfort level 3 and 4.
\newcommand{\tablePercCostCmp}{0.145}
\begin{table}[H]
\centering
\caption{Shows the accumulated economical savings estimate for space heating over the test periods.}
\begin{tabular}{p{0.125\linewidth} p{0.19\linewidth} p{\tablePercCostCmp\columnwidth} 
p{\tablePercCostCmp\columnwidth} p{\tablePercCostCmp\columnwidth}}
\toprule[1.5pt]
Comfort level & Average benchmark cost [\currency]  & Exp. cost [\currency] & Reduction [\currency] & \SavingRateText\ [\%]\\ 
\midrule[1.5pt]
1 ({\color{ref1}\FilledBigTriangleUp}) & 10.92 &  7.33 & 3.59  &32.8 \\ 
2 ({\color{ref2}\FilledBigTriangleUp})& 49.84 & 35.34 & 14.50  &29.1 \\ 
3 ({\color{ref3}\FilledBigTriangleUp}) & 126.42 & 123.49 & 2.93  &2.3 \\ 
4 ({\color{ref4}\FilledBigTriangleUp}) & 42.65 & 35.23 & 7.42  &17.4 \\ 
\midrule[1.5pt]
3 and 4 & 169.07 & 158.72 & 10.35  &6.1 \\
\midrule[1.5pt]
All & 229.83  & 201.39  & 28.43  & 12.4 \\
\bottomrule[1.5pt]
\end{tabular}
\label{table_cmp_cost}
\end{table}
Table \ref{table_cmp_cost} shows a significant \percSaving\ percentage point saving on the heating bill, but looking at the absolute savings of \currency\totalSavings, derived from \numTestDays\ operation days, the results are more modest. Further, it can be seen that the comfort level has a significant impact on savings. \figref{fig_savings_development} shows the development of the estimated \savingRateText\ for each comfort level. The large variance in daily savings means that the long-term expected \savingRateText\ has not settled after 10 days.
\figref{fig_cost_compare} contains the cost results broken down into individual test days (One test day per column), which are analysed with respect to average ambient temperature and sun intensity. Dots are similar benchmark days derived according to description of section \ref{sec_validation} and the black lines in each column represents the average cost of similar benchmark days. The results reveal three main ambient temperature regions: the warm (6 to 13 \si{\degreeCelsius}), the medium (0 to 6 \si{\degreeCelsius}) and the cold (-5 to 0 \si{\degreeCelsius}). In the warm region, the heating demand is so low that percentage losses or gains amounts to very small differences in savings or losses. The medium region shows the highest potential for savings. 
The results from cold, sunny days are difficult to assess due to a sparse amount of similar days present in the benchmark data, but the immediate results point at consistent losses. Further, it can be seen that sunny days (reddish dots and crosses) reduce costs since they drop lower than their more cloudy counterparts. This is of course related to overheating events, which might be uncomfortable. The plot also shows that the costs increase as temperature decreases.

\begin{figure}[H]
	\centering
	\includegraphics[width=1\columnwidth]{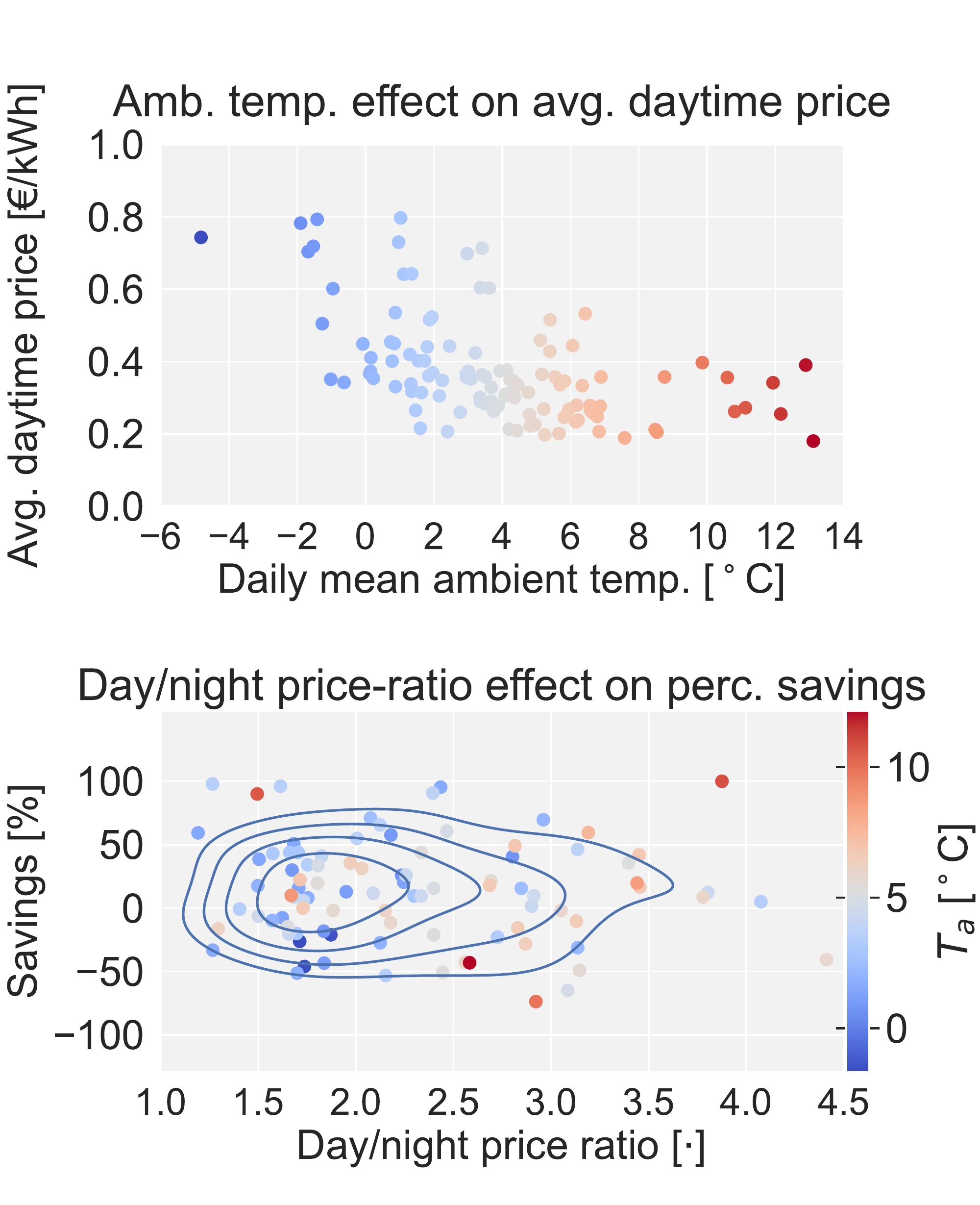}
	\caption{\figUpper Average daytime price as a function of the average temperature. \figLower Daily savings formulated in percentage as a function of the day over night price ratio. Colors indicate the daily average ambient temperature and the contours the density.}
	\label{fig_savings_analysis}
\end{figure}

Table \ref{table_cmp_electricity} shows the electricity consumption. It is common that studies experience an increase in primary energy consumption when applying price responsive control (15.8\% more electricity in \cite{hu_price-responsive_2019}, 10.3\% more heat in \cite{amato_dual-zone_2023}), as is the case for comfort level 1 and 4, albeit the values observed here are significantly higher. Comfort level 2 and 4 show a reduction, which is likely to be connected to the lower indoor temperature in comfort level 1 and a sequence of sunny days which the MPC controller could capitalize on.
\renewcommand{\arraystretch}{1.5}
\newcommand{\tablePercElectricity}{0.23}
\begin{table}[H]
\centering
\caption{Accumulated electricity consumption from the experiment and benchmark data}
\begin{tabular}{p{0.13\linewidth} p{\tablePercElectricity\linewidth} p{\tablePercElectricity\columnwidth} 
p{0.20\columnwidth}}
\toprule[1.5pt]
Comfort level & Accumulated average electricity (Benchmark) [\si{kWh}] & Accumulated Electricity
(Experiment) [\si{kWh}]   & Percentage increase [\%]\\
\midrule[1.5pt]
1 ({\color{ref1}\FilledBigTriangleUp})  &  41.4 & 45.9 &11.0 \\ 
2 ({\color{ref2}\FilledBigTriangleUp}) &  110.5 & 96.3  &-12.8 \\ 
3 ({\color{ref3}\FilledBigTriangleUp}) &  398.9 & 492.9  &23.6 \\ 
4 ({\color{ref4}\FilledBigTriangleUp})  &  191.8 & 178.0  &-7.2 \\ 
\midrule[1.5pt]
3 and 4 & 590.72 & 670.96  & 13.6 \\ 
\midrule[1.5pt]
All   & 743 & 813 & 9.5 \\
\bottomrule[1.5pt]
\end{tabular}
\label{table_cmp_electricity}
\end{table}
As with electricity, the heat production (Table \ref{table_cmp_energy})  has increased, but percentage-wise, not as much. This can be explained by the lower \cop\ causing less heat to be produced for the electricity.
\newcommand{\tablePercEnergy}{0.23}
\begin{table}[H]
\centering
\caption{Accumulated heat produced by the \hp.}
\begin{tabular}{p{0.13\columnwidth} p{\tablePercEnergy\columnwidth} p{\tablePercEnergy\columnwidth} 
p{0.2\columnwidth}}
\toprule[1.5pt]
Comfort level & Accumulated Average heat (Benchmark) [\si{kWh}] & Accumulated heat (Experiment) [\si{kWh}]   & Percentage Increase [\%]\\
\midrule[1.5pt]
1 ({\color{ref1}\FilledBigTriangleUp})  &  192.6 & 218.3 &13.4 \\ 
2 ({\color{ref2}\FilledBigTriangleUp}) &  455.1 & 385.7  &-15.2 \\ 
3 ({\color{ref3}\FilledBigTriangleUp}) &  1667.2 & 1919.0  &15.1 \\ 
4 ({\color{ref4}\FilledBigTriangleUp})  &  799.9 & 707.5  &-11.5 \\ 
\midrule[1.5pt]
3 and 4 & 2467.13 & 2626.56  & 6.5 \\ 
\midrule[1.5pt]
All  & 3115  & 3231 & 3.7 \\
\bottomrule[1.5pt]
\end{tabular}
\label{table_cmp_energy}
\end{table}
\figref{fig_cost_compare} gives a sense of monetary savings potential, but it hides some important factors leading to significant savings larger than \euro 1. \figref{fig_savings_analysis} explores these factors. 
The upper figure clearly shows an inverse correlation between daily average ambient temperature and the average daytime electricity price during this heating season. This partly explains the larger savings potential between 1 and \SI{5}{\degreeCelsius} seen in \figref{fig_cost_compare}. The price is in the high end while the house still maintains flexibility. The next influential factor is volatility in prices which is explored in the lower graph of \figref{fig_savings_analysis}. Here the daily relative savings are plotted against the day over night average price ratio. Nighttime price is defined as the average price between 00:00 and 06:00, and daytime is given as the average over the remaining period. Although the dots are more scattered here, some important trends can be seen. First, the price ratio decreases with colder temperatures. Second, most days with a price ratio above three resulted in savings and, third, cold days gave significant loses.

%% file: sections/results/heat_pump_efficiency.tex
\subsection{Heat pump efficiency model}
\label{sec_heat_pump_efficiency}
\begin{figure}[H]
	\centering
	\includegraphics[width=\columnwidth]{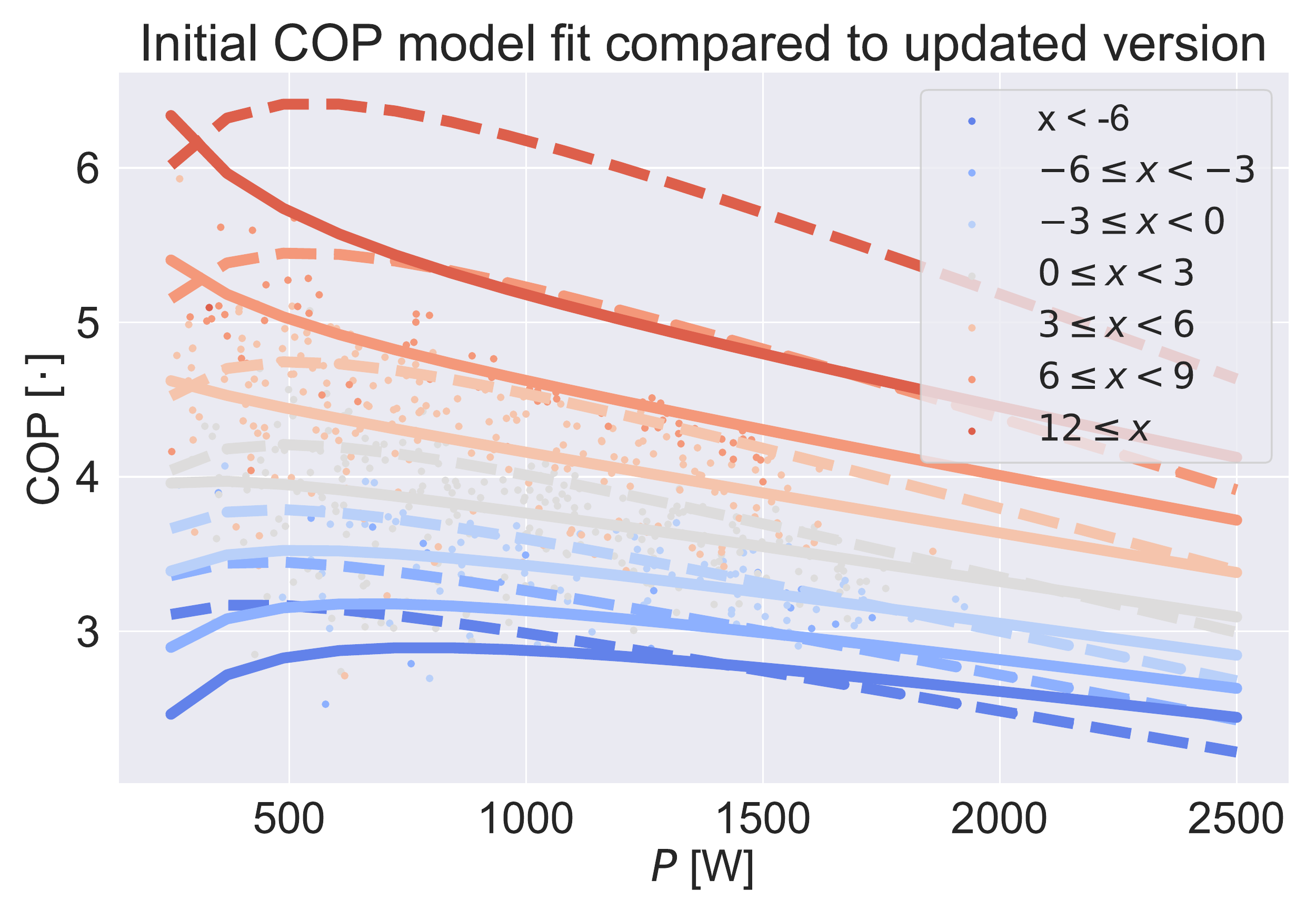}
	\caption{The two parameter fits for the \hp\ efficiency model during the experiment. The dashed lines describe the \cop\ as a function of electricity consumption at varies fixed ambient temperatures. The solid lines show the updated fit, which matches the data (scatter dots), obtained during the experiment, better than the original. The coloring indicates the ambient temperature level with blue signaling cold and red warm.}
	\label{fig_cop_investigation_2}
\end{figure}

\begin{figure}[H]
	\centering
	\includegraphics[width=\columnwidth]{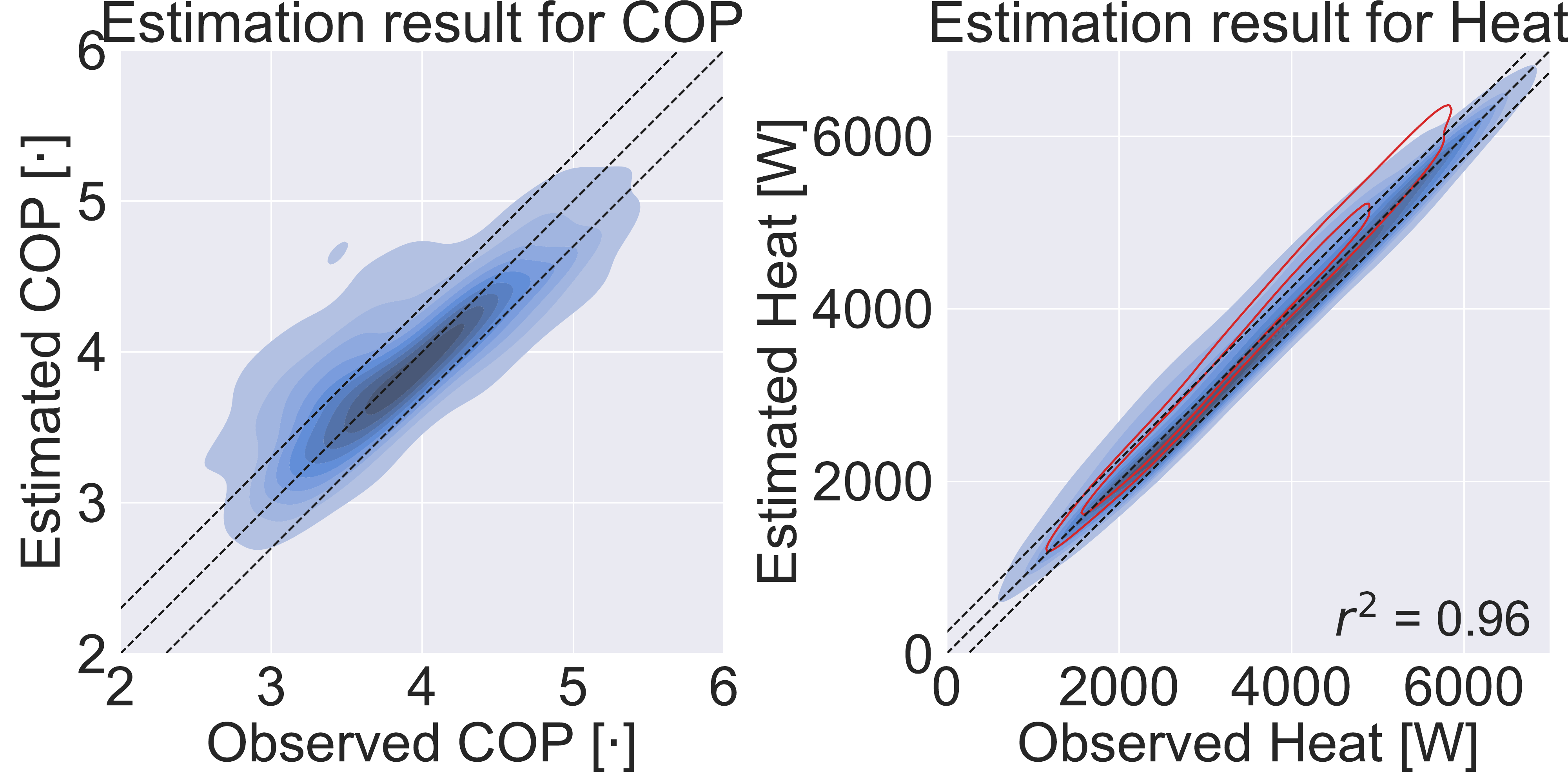}
	\caption{The red contour lines shows the distribution of the new data calculated using the pre-experiment fit.}
	\label{fig_cop_meas_est}
\end{figure}

As the \hp\ efficiency model informs the \MPC\ on the trade-off between heat boosting and efficiency loss, it is highly important that it is accurate. Figure \ref{fig_cop_investigation_2} shows that the general \cop\ fell when the \hp\ was operated according to the new controller, leading to inaccurate estimates.
The dashed lines represents the original fit, which is based on data from the benchmark controller at various fixed ambient temperatures, the scattered dots represents data points obtained in the test period and the solid lines are from an updated parameter fit. The original fit has an $R^2$ value of \rsquaredInit, but since it performed poorly during operation---often overestimating the efficiency---the fit was updated (solid lines), which resulted in lower predicted efficiencies. The main take away is that it is necessary to update the model repeatedly when the operational style changes, otherwise severe miscalculations are introduced. \figref{fig_cop_meas_est} present the results of the updated fit.

%% file: sections/results/production.tex
\subsection{Space-heating production patterns}
\label{sec_production}
\begin{figure}[hbt!]
	\centering
	\includegraphics[width=\columnwidth]{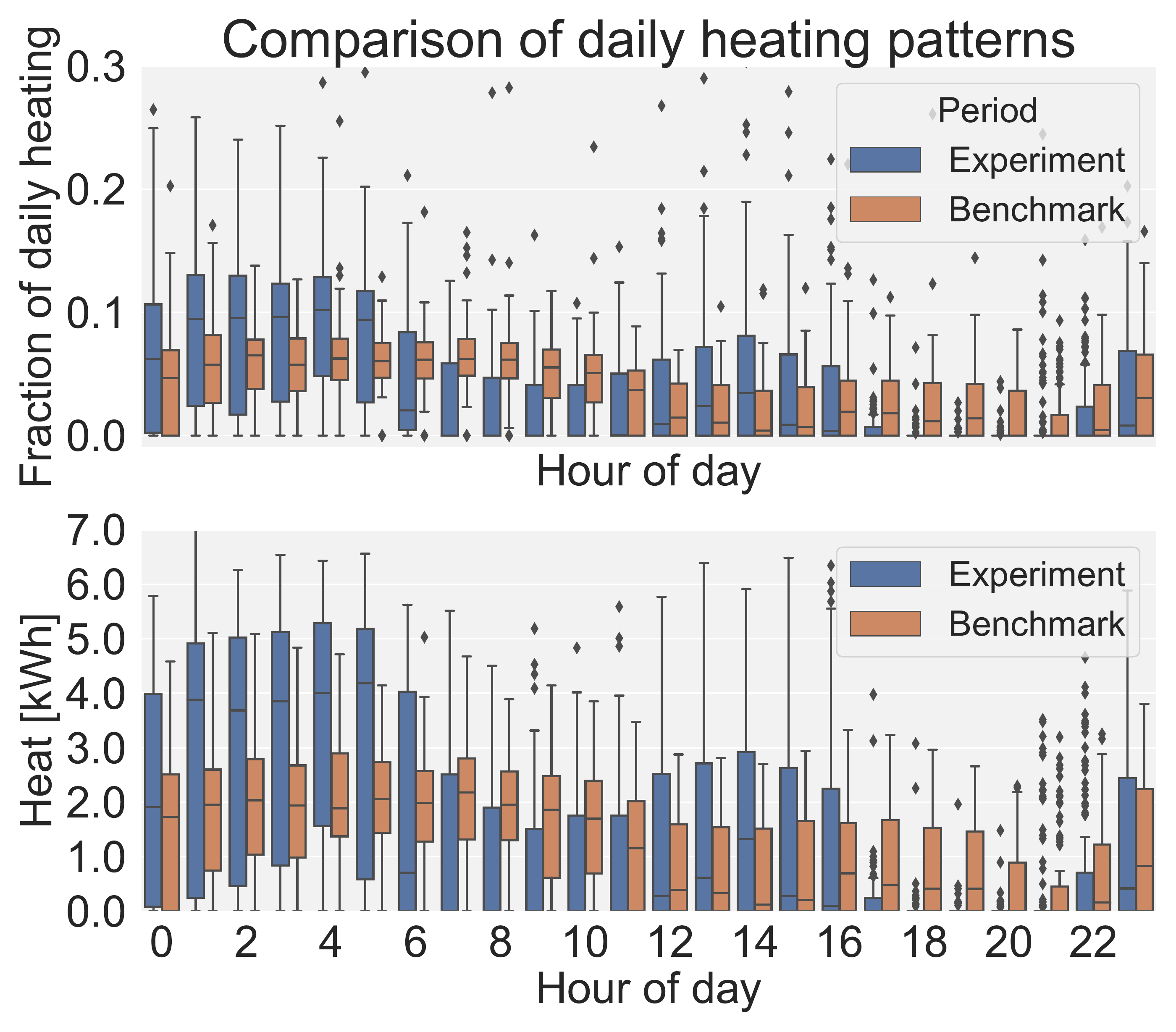}
	\caption{\figUpper\ Distribution of daily production patterns from the benchmark controller compared with the \MPC\ controller. The y-axis is the hourly fraction of total production for a given day. \figLower\ Actual hourly heat production}
	\label{fig_production}
\end{figure}

In this section the daily heat production patterns are presented and compared to the benchmark data. The upper graph in \figref{fig_production} shows the daily heat production curves normalized with respect to part of the full day. Lower shows the actual hourly production in kWh.
The heat production has increased remarkably at night meaning that the controller bets against the classic night setback strategy despite it generally being colder at night causing the \hp\ to be less efficient. The midday production has increased, but most notably is the complete lack of heating in the \cookingPeak\ period between 17:00 and 21:00.

%% file: sections/conclusion/conclusion.tex
\input{sections/conclusion/interpretation.tex}
\input{sections/conclusion/discussion.tex}

\section{Conclusion}
\label{sec_conclusion}
During this study an implementation oriented, price-responsive \MPC\ controller has been tested on a commercial \hp, over the course of four months in the winter 2022-2023. The results show that \loadshifting\ can reduce heating costs by at least \comfThreeCostPerc\%, only by activating the heat capacity of the building structure and without reducing the indoor temperature. The production patterns have been shifted to support the grid through increased consumption at night and by shutting the \hp\ down in the \cookingPeak. Further, it has been established that under the current danish price scheme the \cookingPeak\ is the decisive cost factor, and about \percPeakSave\% of the savings provided by the \MPC\ can be obtained just by blocking the \hp\ in the \cookingPeak. Full or partial shutdown in the \cookingPeak should immediately be broadly implemented. This rule creates correlated consumption patterns, which might become problematic for the grid later. In case the grid operators wish to use more coordinated approach controllers of the type presented here are needed, but, at the moment the cost reduction obtained from price responsiveness cannot cover the costs of acquiring such capabilities, so more financial incentive needs to be provided.

The ambient temperature overwrite applied to control the heat flow of the \hp\ has proven to be a functional but inefficient way to make the it \smartGrid-ready. A dedicated input for reference control as a standard is to be desired if advanced control of \hp s should be the norm. 

Several publications have suggested that the upper layer, in a hierarchical control structure can be controlled using a building model having only one heating zone without degrading indoor comfort. We can report that the results presented here support this idea. Although, it has to be mentioned that the highly insulated shell of the house might me a large contributor.

Future work is to automate the process of gathering quality data from the sensors and apply an update the models for building, \hp\ and \pv\ regularly. The next natural step for the \MPC\ is to upgrade the \hp\ retrofit to include control of domestic hot water production which, at this moment, is a randomly occurring process, often taking place in the \cookingPeak.

%% file: sections/conclusion/interpretation.tex
\section{Interpretation of the results}
\label{sec_result_interpret}
In this section the authors provide their interpretation of the data and results presented in the former section. Starting with Table \ref{table_cmp_cost}, where comfort levels 1 and 2 (\comfLvlTriangleOne\ and \comfLvlTriangleTwo, respectively) show a clear percentage-vise savings potential. At comfort level 1, the indoor temperature was uncomfortably low, so this result is ignored. Test period 2 (\comfLvlTriangleTwo) is more interesting since the residents experienced a comfortable indoor climate while saving on heating costs. This raises the question: \textit{Did the price responsiveness cause the economical
savings?} The answer is unknown since the lower average indoor temperature, and thereby lower heat demand, could have been the main reason. The main take-away from comfort level 2 is that even a \NZEB\  can gain by lowering indoor temperature. Test period three (\comfLvlTriangleThree) was executed with an average indoor temperature of \SI{0.2}{\degC} higher than the one of the benchmark data, meaning that the \comfThreePercSave\% savings are likely to be contributed to the controller.

Having established that overall savings are possible under the current Danish price scheme, the next part investigates situations which are particularly favorable or unfavorable for the controller. 
Before reading on, keep in mind that \textit{large savings can only originate from situations with large potential costs}. For the analysis Figs.~\ref{fig_avg_temp_1}, \ref{fig_cost_compare}, \ref{fig_savings_analysis}, \ref{fig_cop_investigation_2}, Table \ref{table_cmp_electricity} and \ref{table_cmp_energy} are used. \figref{fig_cost_compare} reveals that the largest share of consistent savings are generated between 0.7 and \SI{4.0}{\degreeCelsius}. The large loss seen at \SI{1.1}{\degreeCelsius} is the transition day between test period 2 and 3 where extra energy was needed to lift the average indoor temperature. The lower ambient temperature increases the demand for heat which increases the cost. However, as this is true for all buildings in the region not only the consumption dependent costs are driven up, so are the electricity prices. This is visible in the upper graph of \figref{fig_savings_analysis}, where the average daytime price is inversely correlated with the temperature. The result is that heat demand and price amplifies each other which increase the daily cost dramatically when the ambient temperature is below \SI{5}{\degreeCelsius}.

Having established the factors driving up costs, we turn the attention to daily price variation which also impacts the potential for cost savings, see the lower graph in \figref{fig_savings_analysis}. Although, the results are more scattered than in the upper graph, three trends can be seen. First, the day/night price-ratio is more likely to be higher at high ambient temperatures. Second, at ratios above three, the controller is likely to save money, albeit these are mostly warm low cost days. Third, the most interesting trend is the range 0 to \SI{5}{\degreeCelsius}, where the ratio often was above 2, securing significant percentage-vise savings.

At this point the price conditions for savings are established. Hence, we return focus to test periods 3 and 4 (\comfLvlTriangleThree\ and \comfLvlTriangleFour, respectively) and ask: \textit{Is the potential for savings larger than presented here?} The controller responded to the price signal by changing the daily heating pattern significantly, as seen in \figref{fig_production}. Several things impact savings: model errors, forecasts errors, lack of controller robustness and more. This said, the loss of \hp\ efficiency, mentioned in Section \ref{sec_heat_pump_efficiency}, stands out as a major plausible limiter. 
The efficiency loss originates not only from the higher loads but also due to the dynamic operations. The loss from dynamic operations puts load shifting at a disadvantage. When the supervisory controller calculates the heating plan it also considers the continuous approach featured by the benchmark controller but discards it as inefficient compared to the night heating approach. This happens because the controller relies on the \hp\ efficiency model, which does not inform that the default controller---the one from the manufacturer---can operate the \hp\ more efficiently. This can be used as a critique of the presently implemented heat controller, yet, it can also be posed as a question to why the manufacturers of domestic \hp's do not let them be controlled according to a heat reference as an alternative to the ambient temperature heating curve.

A weakness has been noticed in the \MPC s reliance on forecasts. The procedure has issues dealing with sunny days, even though the predictive nature should ensure superiority of the \MPC. \figref{fig_cost_compare} clearly reveals that days with significant loses tend to have high sun intensity. 
We expect this to be due a combination of several factors which coalesce with unfortunate outcomes. The low electricity prices invite the controller to boost heating intensely between 00:00 and 06:00 to avoid electricity consumption during more expensive day hours. If the model and forecasts were perfect the heat would be boosted accurately. However, in practice an overheating event is likely to occur if a thick cloud cover is wrongfully predicted, and intense sunshine happens instead. The cloud cover data from the weather service has several times been unreliable even at time-of-use. This effect can be mitigated by correcting the forecast with live \pv\ data. Further, a robust control approach which restrains night boosting slightly should be applied.

%% file: sections/conclusion/discussion.tex
\section{Discussion}
\label{sec_discussion}
Having shown a savings potential through price responsive \loadshifting, the following topics deserve attention: the step from simulation to reality, potential performance improvements, control of the \hp, heat scheduling using an average indoor temperature and minimizing operation costs rather than discomfort or indirect \coTwo-emissions.

A large amount of papers assume perfect forecasting when conducting simulation studies of \MPC\ with the consequence that results reflect upper performance boundaries. This is avoided in a real implementation. Nevertheless, the problem then shifts to estimating the true cost reduction or \savingRateText. \figref{fig_savings_development} shows the extent of the challenge since the \savingRateText\ has not converged after 10-12 days. Even after 55 days this is not fully the case. The reason is the high volatility in daily savings and losses which are in the range $\pm$\euro3. The implications are that short-term studies (of the order of days) are at risk of reporting \savingRateText s which diverge severely from the true rate. If the so called ``File Drawer Effect'' (Failing to publish negative results) is at play, the bias might be towards too high savings potential. The strategy applied here is to rely on benchmark data collected from the prior heating season. However, even with a full season of data, there are holes in the coverage, meaning that there are experiment days without counterparts in the benchmark dataset. Ideally, the sensing equipment needs to be installed several seasons before the experimental controller is applied in order to have a reliable dataset.

The main focus areas for improved performance are \hp\ control and modelling. The performance of the \MPC\ was degraded by problems listed below.
\input{sections/results/lessons}
All of these effects could have been prevented if the \hp\ had an interface for set-point control. The path forward for commercial heat pumps should be to provide an interface for reference control which would allow the \hp\ to operate in a near-optimal state while being part of a coordinated and cooperative control scheme.

Using an averaged room temperature in the upper control level has proven to be completely viable with respect to comfort. Controlling this way does conflict with the idea that each room should be controlled strictly after individual references, but it is our experience that large temperature differences within the thick shell of a low energy house are difficult to obtain in any case.

Given that prices are the result of market mechanisms rather than purely physical processes, it is exceedingly difficult to forecast future prices. This controller can become more efficient if the daily price-volatility increases, on the other hand, lower prices can make the controller superfluous. This said, low prices are good for the consumer, so one might think of the controller as an insurance against long periods of high, volatile prices.

During the experiment period, the models were updated several times with the latest data. Although This was done manually, it is not our recommendation to implement a similar control structure without a proper automated procedure in place. It is simply too costly to perform updates manually in a real setting.

Lastly, a look at some economic aspects with respect to the current price situation and controller performance. If the heating season lasts 4 months ($\approx$ 120 days), an average cost reduction of \currency1 per day would translate into \currency1200 over a 10 year period.Although it is hard to predict, this could be a reasonable price for the controller. The \MPC\ yielded a reduction of \currency\savingRate\ per day, meaning that it would take \yearsEarning\ years to save \currency1200 under current price conditions. It has to be noted that this case involves a low energy house, and this calculation is not meant to be extrapolated to less energy efficient houses. 

The analysis of the savings potential for the \MPC\ controller would be incomplete if the effect of the relatively large \cookingPeak\ price is ignored. The daily recurrence of the \cookingPeak\ tariff begs the question: \textit{What is the \savingRateText\ from simply blocking the \hp\ in the \cookingPeak\ period?} Using the method from section \ref{sec_validation}, the benchmark controller has used an extra \extraPeakElec\ \kwh\ electricity in the timespan 17:00-21:00 translating to an extra cost of \currency\extraPeakCost\ over the test period compared to the \MPC\ approach. Postponing the electricity consumption to the hours following the \cookingPeak\ would cost \currency\costAfter, based on the average price between 21:00 and 01:00, resulting in an overall reduction of \currency\savePeakBlock. This is \percPeakSave\% of the estimated savings provided by the \MPC. Two things have to be noted, the peak block saving assumes a \cop\ of 4.2, which can only be achieved at moderate heating loads, and the cost reduction of the experiment is calculated based on all comfort levels.

%% file: sections/results/lessons.tex

\begin{itemize}
    \item A side effect of using the compressor block function is that the \hp\ attempts to heat the \DHW\ using the electric heating rod, which has a power output of \SI{10}{kW}. This is far from ideal, since just a few minutes in this state is costly. \textbf{Suggested solution}: block the compressor for space heating only.
    \item When the \hp\ defrosts, the measured heat flow reverses. Any controller regulating the heat output needs to be able to detect and handle such a situation. \textbf{Suggested solution}: put  the control in standby mode.
    \item It is not possible to start the heat pump on demand, the only option is to release the compressor brake and wait. The waiting time is observed to be between 60 and 120 min. \textbf{Suggested solution}: adapt the block release for best start-up timing or introduce/use open HP controller standards
    \item In certain situations the heat pump shuts down before it should. It is assumed that a combination of high ambient temperature and low flow caused the internal controllers to shut it down. \textbf{Suggested solution}: use data to figure out what events cause a shut down.
    \item A low pass filter and other unknown internal states make control through ambient temperature overwrite particularly challenging. \textbf{Suggested solution}: use a heat pump with reference control for heat.
\end{itemize}





%% file: sections/admin/admin.tex
\section*{Credit author statement}
\textbf{Simon Thorsteinsson:} Conceptualization, Methodology, Software, Formal analysis, Investigation, Data Curation, Writing - Original Draft, Visualization \textbf{Alex Kalaee:} Conceptualization, Writing - Review \& Editing \textbf{Pierre Vogler-Finck:} Conceptualization, Writing - Review \& Editing \textbf{Henrik Stærmose:} Resources, Conceptualization, Writing - Review \& Editing, Funding acquisition \textbf{Ivan Katic:} Conceptualization, Writing - Review \& Editing, Project administration, Funding acquisition  \textbf{Jan Bendtsen:} Conceptualization, Resources, Writing - Review \& Editing, Supervision, Funding acquisition.
\section*{Declaration of Competing Interest}
The authors have no competing interests to declare.

%% file: sections/appendix/appendix.tex
\input{sections/appendix/resident_statement.tex}
\input{sections/appendix/app_model_fits.tex}

%% file: sections/appendix/resident_statement.tex
\section{Resident statements}
\label{app_resident_statement}
The $11^\text{th}$ Jan. 2023, the residents were sent a questionnaire about the experienced indoor climate comfort. Before reading the questionnaire, note that it was not conducted anonymously and the text is translated from Danish. The questions and answers are as follows.
\begin{enumerate}
    \item \textit{Have you experienced an increase in discomfort with respect to indoor climate when you compare with last heating season?}
    \begin{description}
    \item [Answer:] I actually think that we have felt a better comfort than I remember from last year. There was an experience of discomfort in the beginning of the experiment, which we talked about, thereafter it has been quite comfortable.
    \end{description}
    \item \textit{When you have experienced discomfort, has it been too warm, too cold or do you experience both too warm and too cold periods?}
    \begin{description}
    \item [Answer:] No, we don't have any periods with unpleasantly low temperatures. We have as always lower temperatures in the cinema/hobby-room, but that has been fine.
    \end{description}
    \item \textit{Is there any time of day where the discomfort most often occurs?}
    \begin{description}
    \item [Answer:] No comment
    \end{description}
    \item What have you noticed with respect to floor temperatures?
    \begin{description}
    \item [Answer:] It has actually been pleasantly warm, and I don't think that we have experienced cold floors, which we typically experience at middle-high outdoor temperatures.
    \end{description}
    \item Have you experienced that the radiation from the floors has been too high?
    \begin{description}
    \item [Answer:] No
    \end{description}
    \item Have you experienced that the radiation was too low? A feeling of being cold even though the room temperature was high.
    \begin{description}
    \item [Answer:] No, as said, we have not experienced that for long.
    \end{description}
\end{enumerate}

%% file: sections/appendix/app_model_fits.tex
\section{Model fits}
\input{sections/results/cop_regr}

\input{sections/results/house_model_param}

%% file: sections/results/cop_regr.tex
\subsection{Heat pump efficiency model}
\begin{figure}[H]
	\centering
	\includegraphics[width=1\columnwidth]{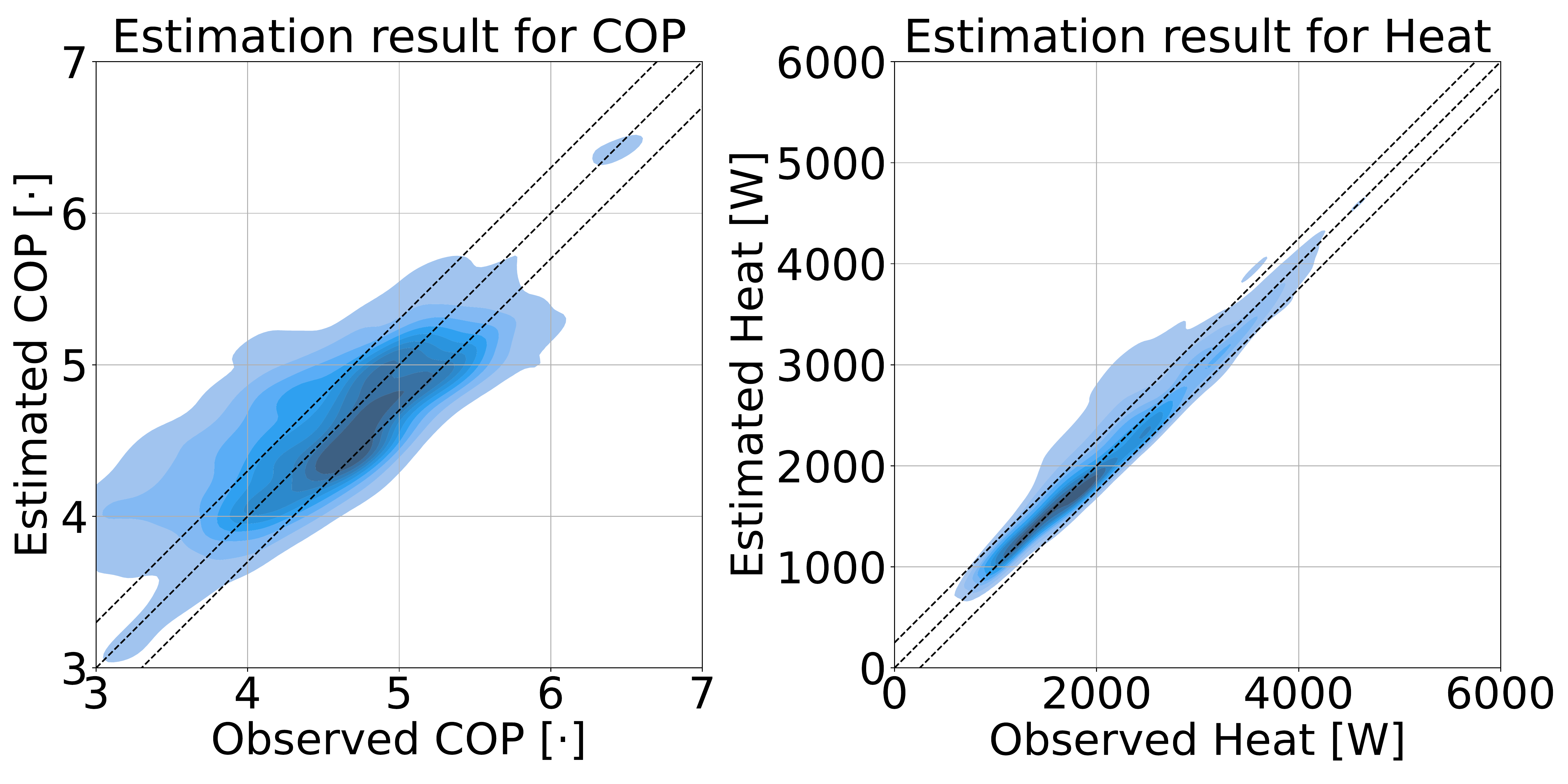}
	\caption{A diagram of the heat pump combined with the floor heating system.}
 	\label{fig_dQ_fit}
\end{figure}
\newcommand{\tablePerc}{0.12}
\begin{table}[H]
\centering
\caption{Fitted parameters for heat pump efficiency model.}
\begin{tabular}{p{0.18\columnwidth} p{\tablePerc\linewidth} p{\tablePerc\columnwidth} 
p{\tablePerc\columnwidth} p{\tablePerc\columnwidth}  p{0.07\columnwidth}}
\toprule[1.5pt]
date & $k_0$  & $c_0$ & $c_1$ & $c_2$  & $\Bar{\T}_\forward$  \\ 
\midrule[1.5pt]
2022-11-05&125.256 & -25.348 & 414.026 & -62.854 & 26.63\\
2022-11-24&-9288.9 & 2363.53 & 1325.13 & -210.53 & 50.00\\ 
2022-12-01&-1880.2 & 273.2 & 694.6 & -101.28 & 50.00\\ 
2023-01-27&-793.31 & 105.79 & 509.07 & -46.854 & 41.00\\ 
\bottomrule[1.5pt]
\end{tabular}
\label{table:1}
\end{table}
\vspace{\tableSpace}

%% file: sections/results/house_model_param.tex
\subsection{House model}
\label{app_house_model}
This section shows the first and last parameter fit for the house model described in Section \ref{sec_house_model}.
\begin{figure}[H]
	\centering
	\includegraphics[width=1\columnwidth]{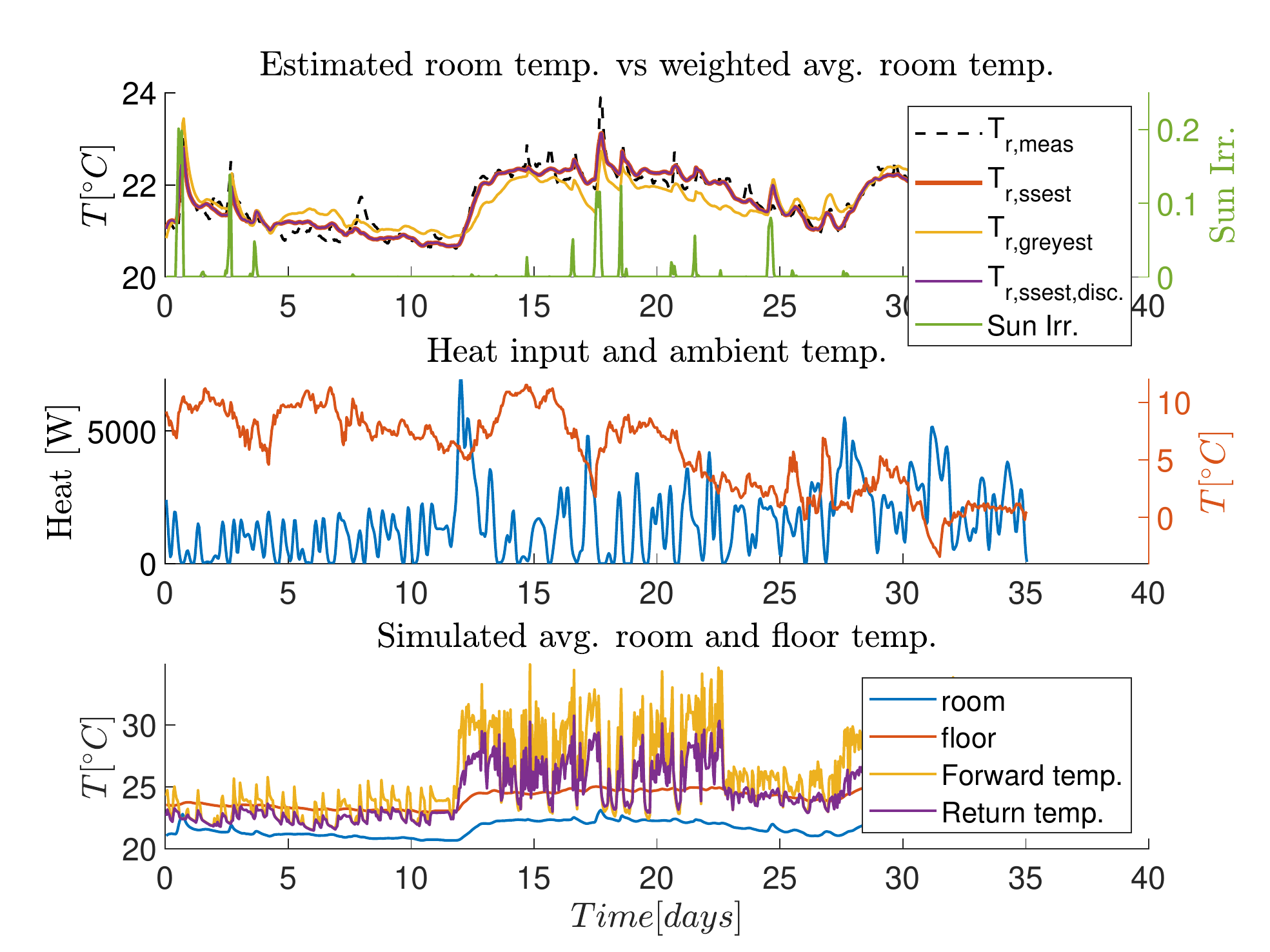}
	\caption{Fitted 2022-02-05 to the 35 prior days of data.}
 	\label{fig_averaged_temp}
\end{figure}
\begin{figure}[H]
	\centering
	\includegraphics[width=1\columnwidth]{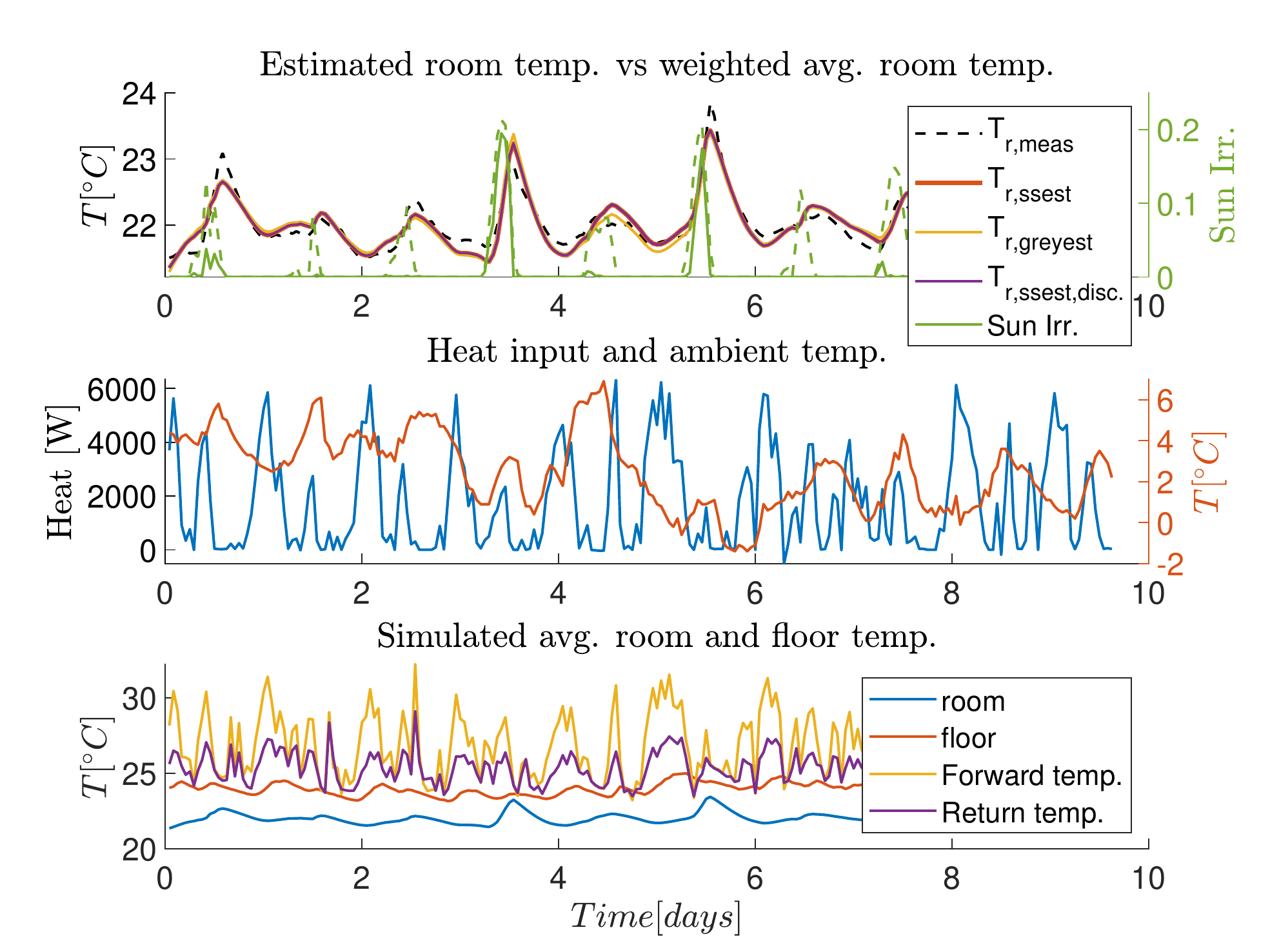}
	\caption{Fitted 2023-02-09 to the 10 prior days of data.}
 	\label{fig_averaged_temp_2023-02-09}
\end{figure}